\documentclass{article}
\usepackage{arxiv}

\usepackage[utf8]{inputenc} 
\usepackage[T1]{fontenc}    
\usepackage{hyperref}       
\usepackage{url}            
\usepackage{booktabs}       
\usepackage{amsfonts}       
\usepackage{nicefrac}       
\usepackage{microtype}      
\usepackage{lipsum}
\usepackage{graphicx}
\graphicspath{ {./images/} }

\usepackage{amsmath}
\usepackage{xcolor}
\usepackage{colortbl}
\usepackage{float}
\usepackage{ragged2e}

\definecolor{excelgreen}{rgb}{0, 0.69, 0.31} 
\definecolor{lightgreen}{rgb}{0.776, 0.937, 0.808}
\definecolor{lightgray}{rgb}{0.851, 0.851, 0.851}

\title{Simplifying complex machine learning by linearly separable network embedding spaces}

\author{
 Alexandros Xenos \\
  Barcelona Supercomputing Center\\
  08034 Barcelona, Spain\\
  \texttt{alexandros.xenos@bsc.es} \\
   \And
 Noel Malod-Dognin \\
  Barcelona Supercomputing Center\\
  Barcelona, Spain, 08034\\
  \texttt{noel.malod@bsc.es} \\
  \And
 Natasa Przulj \\
Barcelona Supercomputing Center\\
08034 Barcelona, Spain\\
Department of Computer Science\\
University College London\\
WC1E 6BT London, UK\\
ICREA, Pg Lluís Companys 23\\
08010 Barcelona, Spain\\
\texttt{natasha@bsc.es}\\
}
  
\date{} 

\begin{document}
\date{} 
\maketitle
\begin{abstract}
Low-dimensional embeddings are a cornerstone in the modelling and analysis of complex networks. However, most existing approaches for mining network embedding spaces rely on computationally intensive machine learning systems to facilitate downstream tasks. In the field of NLP, word embedding spaces capture semantic relationships \textit{linearly}, allowing for information retrieval using \textit{simple linear operations} on word embedding vectors. Here, we demonstrate that there are structural properties of network data that yields this linearity. We show that the more homophilic the network representation, the more linearly separable the corresponding network embedding space, yielding better downstream analysis results. Hence, we introduce novel graphlet-based methods enabling embedding of networks into more linearly separable spaces, allowing for their better mining. Our fundamental insights into the structure of network data that enable their \textit{\textbf{linear}} mining and exploitation enable the ML community to build upon, towards efficiently and explainably mining of the complex network data.
\end{abstract}


\section{Introduction}

Networks naturally model complex systems in many real-world applications. Examples include social, information, and biological domains. Current state-of-the-art approaches for analyzing these complex data are based on network embedding techniques \cite{nelson2019embed}. These algorithms map a network nodes in a low \textit{d}-dimensional space, where the geometry of the space reflects the similarities between the nodes \cite{li2022graph} in the sense that two nodes are defined to be \emph{similar} either when they belong to the same network neighbourhood, or have similar topological roles independent of being adjacent, e.g. being hub nodes (also called \emph{topological similarity}). In some data types, it is unclear how to exploit the structure of the generated embedding spaces, so the vectorial representation of the nodes (termed \emph{embedding vectors}) are then fed to computationally intensive machine learning (ML) systems to perform tasks, such as node classification, clustering, and link prediction.  Interestingly, in another domain, in the field of Natural Language Processing (NLP), the Skip-Gram neural network (NN) based word embeddings (e.g. \textit{Word2vec} \cite{mikolov2013efficient}) were shown to capture semantic relationships between words linearly, allowing for downstream analysis tasks using simple linear operations on word embedding vectors.\cite{mikolov2013efficient} More recently, linear semantic relationships (i.e., the compositionality of embedding vectors) have also been observed in data embeddings from pre-trained vision-language models (VLMs).\cite{trager2023linear} This poses the question of why in some cases, the embedding techniques lead to a well-stratified embedding space amenable to linear exploitation, while in other cases they do not. This motivates us to explore: (i) if there is an intrinsic property in the structure of the data that yields this linearity and (ii) if the linearity holds in the biological network domain, for embeddings of the systems-level molecular networks. If so, this would simplify the analyses by making them linear, hence alleviating the need for computationally intensive ML models.

The main methods for generating network embeddings are using Graph Neural Networks (GNNs) that adapt the Graph Convolutional Networks used in image processing \cite{kipf2016semi,velivckovic2017graph}, or employ random walks to generate sequences of nodes (e.g., \textit{DeepWalk} \cite{Perozzi2014}, \textit{LINE} \cite{Tang2015} and \textit{node2vec} \cite{Grover2016}) on which the above mentioned Skip-Gram NN architecture  \cite{mikolov2013efficient} from NLP for word embeddings is then applied. Interestingly, Qiu \textit{et al.} \cite{Qiu2018} showed that such Skip-Gram based approaches are implicitly factorizing random-walk-based mutual information matrices, for which they proposed a closed formula. Importantly, they showed that they could obtain equivalent or better network embeddings for downstream analysis tasks by directly factorizing these matrices using Singular Value Decomposition (SVD). Subsequently, we demonstrated that by factorizing these matrices with Non-Negative Matrix Tri-Factorization (NMTF), a popular matrix factorization approach used notably for integrating (fusing) and mining networks, we could achieve equivalent or better results in node classification tasks in biological networks by using simple linear operations on the node embedding vectors.\cite{xenos2021linear} Matrix factorization approaches, including NMF and NMTF, are explainable methods and for that reason have succesfully been applied to integrate and mine biomedical data\cite{Zitnik2013,gligorijevic2016patient,chalise2017integrative,Vitali2018,Malod-Dognin2019,xenos2023integrated} where explainability and interpretability are essential.\cite{combi2022manifesto} Importantly, matrix factorization approaches have been shown to outperform other integration approaches in the context of precision medicine.\cite{cantini2021benchmarking}

Network embedding is challenging, as it involves capturing both structural (topological) and semantic information of a graph (i.e., node labels) \cite{li2022graph}. Based on the distribution of the node labels, the networks are characterized as \emph{homophilic} when nodes with similar labels tend to be adjacent in the network and \emph{heterophilic} otherwise. Note that heterophilic graphs are a data type on which GNNs perform poorly \cite{alon2020bottleneck,zhu2020beyond}. Their poor performance is attributed to the objective function used in most of the network embedding methods, which aims to preserve mainly the neighborhood-based similarity of the nodes. In particular, GNN architectures are learning the vectorial representation of the nodes by aggregating their neighbour information (the so-called message-passing mechanism), while random-walk-based approaches do it by aggregating information from a small context window. Hence, these approaches are more suitable for homophilic networks in which similarly annotated nodes are connected and thus, more likely to appear in the same random walks.

In network biology, it has been shown that genes (nodes) that have similar functions (i.e., similar annotations) can be in different and far away network neighborhoods, but have similar local wiring patterns, i.e. they can be topologically similar \cite{milenkovic2008uncovering,Malod-Dognin2019,xenos2021linear}. The methods that quantify the local wiring pattern similarities are based on graph substructures, which have been recognized as essential to mining complex networks, e.g., cliques in Protein-Protein Interaction (PPI) networks, or triangle counts in social networks \cite{leskovec2008microscopic,bianconi2014triadic}. Such substructures are best captured by \emph{graphlets} \cite{prvzulj2004modeling}, which are small, connected, non-isomorphic, induced subgraphs of a network. Graphlets have been leveraged to generate embeddings that capture the topological similarity of nodes \cite{dutta2018stochastic,rossi2018hone} In our recent work, we used graphlet-based measures to perform random walks between similarly wired nodes, independent of the nodes being adjacent \cite{xenos2021linear}, which enabled us to identify novel human cancer-related genes by performing linear operations on the gene embedding vectors in the PPI network embedding space. To sum up, random-walk based embeddings preserve the neighborhood-based similarity of the nodes and the graphlet-based embeddings preserve their topological similarity. Note that for heterophilic networks,  the latter could be more suitable. Furthermore, depending on the type of similarity being preserved in the embedding space, we can perform different downstream analysis tasks (e.g., predict protein / gene function, or identify cancer-related genes) by applying simple linear operations on the node embedding vectors. This raises the question of whether leveraging both network neighborhood and topological similarities could yield embedding spaces that enable versatile downstream analysis tasks to be performed by simple linear operations on the node embedding vectors.


\subsection{Contribution}

In this study, we introduce novel, graphlet-based, random-walk matrix representations to be used for embedding networks that account for both network neighborhood similarity and topological similarity of nodes. The abundance of these network matrix representations allows us to explore the link between the linearity of the resulting embedding space and the intrinsic properties of the given matrix representation. We hypothesize that the more homophilic the input network matrix representation, the more linearly separable (and hence linearly exploitable) the resulting embedding space, hence diminishing the need for complex ML approaches to perform downstream analyses. We demonstrate that for the thirteen networks from multiple domains that we studied (six multi-label biological networks and seven single-label networks from social, citation, and transportation networks domain), there always exists a graphlet-based matrix representation that yields a more homophilic representation compared to the standard adjacency matrix.  We generate the embeddings by explicitly factorizing these matrices using an Orthogonal NMTF (ONMTF) framework.\cite{ding2006orthogonal} We demonstrate that in 9 out of the 13 embedding spaces we can achieve as good node classification F1-scores with linear classifiers (e.g., linear Support Vector Machines (SVM)) as with non-linear classifiers, indicating that the embedding vectors of nodes from different classes are sufficiently linearly separable. We also demonstrate that in three of these networks, the embedding spaces are fully linearly separable, since the node classification F1-scores of linear SVM are very high and not statistically significantly different than those of complex non-linear ML methods. Finally, we also demonstrate that even when the resulting embedding spaces are not linear (i.e. when the non-linear classifier better disentangles the different node classes), our graphlet-based embeddings outperform the state-of-the-art random-walk based embedding methods, such as as LINE and DeepWalk, by at least 8 \% in the node classification F1-scores. In conclusion, our results demonstrate that our new graphlet-based methodologies for embedding networks into linearly separable spaces allow for higher quality and more efficient mining of networks, also alleviating the need for complex and computationally expensive machine learning methods.

\section{Methods}

\subsection{Datasets}

\paragraph{Biological multi-labeled networks:} In this study, we focus on the most well-studied molecular networks, modelling protein-protein interactions and gene co-expressions of the following three species: \emph{Homo sapiens} (human), \emph{Saccharomyces cerevisiae} (budding yeast) and \emph{Schizosaccharomyces pombe} (fission yeast). For each of these three organisms, we collected the experimentally validated protein-protein interactions (PPIs) from BioGRID \cite{oughtred2019biogrid} version 3.5.182. We model these data as PPI networks in which nodes represent protein-coding genes, and edges connect nodes (genes) whose protein products physically bind.  Also, we collected the gene co-expressions (COEX) from COXPRESdb \cite{coexdb} version 8 (for all the species, we selected the dataset, u22, that has the highest number of expression samples). In our collected co-expression datasets, the co-expression measure between two genes is standardized using zeta scores. To construct the COEX networks, in which nodes represent genes and edges the co-expressions between the genes, we selected the strongest co-expression values having a zeta score higher than or equal to 3, which is the usual practice. The statistics of the six generated biological networks are presented in Table \ref{table:bio_netstats}. These networks are multi-labeled, as a gene often has multiple annotations, detailed below.

\paragraph{Gene annotations:} For each gene (or equivalently, protein, as a gene product) in the biological networks, we collected their Reactome Pathway (RP) annotations from Reactome database v83\cite{fabregat2017} and their KEGG Pathway (KP) annotations from KEGG database v111.1.\cite{kanehisa2000kegg} Also, we collected the most specific experimentally validated Gene Ontology Biological Process (GO BP) annotations \cite{ashburner2000} from the NCBI database (downloaded on September 2023). The GO BP annotations are back-propagated, using the \textit{is-a} relationship, to their ancestors in the Gene Ontology tree. Note that a gene can have multiple GO BP, KP and RP annotations.


\paragraph{Single-labeled networks:} We analyze single-labeled networks, beyond biology, that are used as benchmarking datasets in ML studies. We collected from PyTorch Geometric \cite{Fey/Lenssen/2019} the USA air-traffic network,\cite{ribeiro2017struc2vec} the Coauthor CS network,\cite{shchur2018pitfalls} the CORA and the CiteSeer citation networks,\cite{sen2008collective} the two standard heterophilic Wikipedia page–page networks (Chameleon and Squirrel) \cite{multi-scale} and the Wiki-CS hyperlinks network. \cite{mernyei2020wiki} For all the datasets, we treat the networks as undirected and only consider their largest connected component. The statistics of the aforementioned networks are presented in Table \ref{table:netstats}.


\subsection{Network embeddings}

The state-of-the-art approaches to analyze complex data modeled as networks are by using network embedding algorithms.\cite{nelson2019embed,li2022graph} Such algorithms represent nodes as vectors in a \textit{d}-dimensional space, in which similar nodes are embedded close in the space. The new generation of network embedding algorithms are inspired by the word embedding techniques used in NLP.

Recently,  Levy \textit{et al.} \cite{levy2014neural} showed that neural networks (i.e., Skip-Gram with Negative sampling, SGNS), that are used to obtain the word embeddings in NLP, are implicitly factorizing a word-word matrix in which a cell represents the pointwise mutual information (PMI) \cite{church1990word} of the two words, shifted by a global constant; in particular, PMI measures the strength of the association between two words by calculating the log-likelihood ratio of the co-occurrence of two words, compared to what would be expected if they were statistically independent. However, the PMI value of two words, $w$ and $c$, that do not occur in the lexical corpus is $PMI(w,c) = \log 0 = - \infty .$ To address this issue, in the field of NLP, they replace the negative values with zeros, resulting in a sparser matrix, called the positive PMI (PPMI) matrix.\cite{church1990word} Levy \emph{et al.}  \cite{levy2014neural} also demonstrated that the exact factorization of this PPMI matrix with Singular Value Decomposition (SVD) leads to equivalent, or better results for word similarity tasks than the embeddings resulting from SGNS architectures.


Formally, for two words, \textit{w} and \textit{c}, PPMI is defined as:
\begin{equation}
\label{eq:PPMI_NLP}
PPMI(w,c)= max \left(0, \log \frac{\#(w,c) \times \lvert C \rvert}{\#w \times \#c}\right),
\end{equation}
where $\lvert C \rvert$ is the size of the corpus, $\#(w,c)$ is the number of times the two words co-occur in the corpus and $\#w$ and $\#c$ are the numbers of times the words \textit{w} and \textit{c} occur in the corpus, respectively.

In analogy to this, for network embeddings, Qiu \textit{et al.}~\cite{Qiu2018} showed that the Skip-Gram based network embeddings that rely on random walks, such as DeepWalk \cite{Perozzi2014}, LINE \cite{Tang2015}, PTE \cite{Tang2015a} and node2vec \cite{Grover2016}, are also implicitly factorizing a random-walk-based mutual information matrix \cite{Qiu2018}: this matrix is equivalent to the PPMI matrix from NLP, as its cells quantify how frequently two nodes of the network, \textit{i} and \textit{j}, co-occur in a random walk on a network, compared to what would be expected if the occurrences of the nodes were independent. The matrix closed formula of DeepWalk \cite{Qiu2018} is defined as:

\begin{equation}
\label{eq:Deepwalk}
DeepWalk = max \left(0,\log(vol(A)\left(\frac{1}{T}\sum_{r=1}^{T}(D^{-1}A)^{r}\right)D^{-1})- \log b \right),
\end{equation}
where \textit{vol(A)} is the volume of the network, i.e., it's number of edges and is computed as $vol(A)= \sum_{i} \sum_{j} A_{ij} $, \textit{A} is the adjacency matrix of the network, \textit{D} is the diagonal matrix of degrees of the given network, \textit{T=10} is the length of the random walks and \textit{b} is the negative sampling in Skip-Gram. For computational reasons, in the study of  Qui \textit{et al.}~\cite{Qiu2018}, they set $b=1$ to omit the constant term $-\log b$. They demonstrate that the explicit factorization of the DeepWalk closed formula with SVD leads to equivalent, or better performance in network mining tasks than the implicit Skip-Gram based ones. In our recently published work \cite{xenos2021linear}, we implement the DeepWalk closed formula to compute the PPMI matrix of molecular networks and demonstrate that it leads to better organized embedding spaces compared to the traditionally used Adjacency matrix.

The closed formula of LINE \cite{Qiu2018} is a special case of the DeepWalk closed formula, using a a random walk of length $T=1$. This type of a random walk considers that two nodes can co-occur only if they are directly connected by an edge, and is defined as:
\begin{equation}
\label{eq:LINE}
LINE =  max \left(0, \log\left(vol(A)D^{-1}AD^{-1}\right)-\log b \right),
\end{equation}


\noindent For a specific pair of nodes, $u$ and $v$, the formula is:
\begin{equation}
\label{eq:LINE 2}
LINE(u,v) = max \left(0, \log\frac{vol(A) \times A{uv}}{D_{uu} \times D_{vv}} \right),
\end{equation}
which is the equivalent of the aforementioned PPMI formula, but applied to networks.


In this study, to generate representations that capture simultaneously topological and neighborhood-based similarity, we extend the PPMI and DeepWalk closed formulae by using graphlets as contexts for the random walks, as detailed below.



\subsection{Graphlets and Graphlet Adjacency}
\label{Graphlets}

{\em Graphlets} are small, connected, non-isomorphic, induced sub-graphs of a large network, that appear at any frequency in the network.\cite{prvzulj2004modeling} Different topological positions within graphlets are characterized by different symmetry groups of nodes, called automorphism orbits.\cite{prvzulj2007biological}  Orbits are used to generalize the notion of the node degree: the \emph{graphlet degree} of a node is the number of times the node touches a particular graphlet at a particular orbit.\cite{Milenkovic2008} Yaveroglu \textit{et al.} \cite{Yaveroglu2014} showed that between the orbits, there exist redundancies, as well as dependencies, and proposed a set of 11 non-redundant orbits of 2- to 4-node graphlets (see Figure~\ref{fig:graphlets}). Each node in the network is represented by its 11-dimensional vector called \textit{Graphlet Degree Vector (GDV)}, that captures the 11 non-redundant graphlet degrees of the node. To quantify the topological similarity between two nodes, \textit{u} and \textit{v}, we compare their GDV vectors using the GDV distance, which is computed as follows. Given two GDV vectors, \textit{x} and \textit{y}, the distance between their $i^{th}$ coordinate is defined as: 

\begin{equation}
Dist_{i}(x,y)=w_i \times \frac{\log(x_i +1) - \log(y_i +1)}{\log(max\{x_i,y_i\}+2)},
\end{equation}
where $w_{i}$ is the weight of orbit \textit{i} that accounts for dependencies between the orbits (detailed by Milenkovic and Przulj \cite{Milenkovic2008}). The log-scale is used to control the different orders of magnitude between orbit counts.
Then, GDV similarity between nodes \textit{u} and \textit{v} is defined as:
\begin{equation}
GDVsim(u,v)=1 - \frac{\sum_{i=1}^{11}Dist_i(x,y)}{\sum_{i=1}^{11}w_i}.
\end{equation}
The pairwise GDV similarities over all nodes in a network are represented in the GDV similarity matrix, capturing the topological similarities over all nodes in the network.\cite{Yaveroglu2014} In our recent study, we used the GDV similarity matrix as the input into the DeepWalk closed formula and factorized the resulting matrix (GDV PPMI matrix) with NMTF to generate the network embedding based solely on the topological similarities between the nodes.\cite{xenos2021linear}

The above definitions of the graphlet-based measures quantify the similarity in local wiring patterns around network nodes regardless of them being adjacent, or close in the network. These measures do not consider whether two nodes participate in the same local network neighborhood, i.e., in the same graphlet. To bridge this gap, Windels \textit{et al.}~\cite{graphletlaplacians} introduced the concept of the \emph{Graphlet Adjacency}, in which a pair of nodes is adjacent if they simultaneously touch (participate in) a given graphlet (see Figure~\ref{fig:graphlets} for an illustration of graphlets). This results in Graphlet Adjacency matrices corresponding to each graphlet; e.g., $G_0$ corresponds to the original adjacency matrix and for larger graphlets, two nodes can be adjacent more than once, e.g. between two nodes there may exist more than one 3-node path. Given this extended definition of adjacency, for each graphlet, $G_k$, the corresponding graphlet adjacency matrix, $A_k$, is defined as:

\begin{equation} 
A_k(u,v) = \begin{cases} 
\mbox{$a_{uv}^k$,} & \mbox{if } u \neq v \\ \mbox{0,} & \mbox{otherwise,} 
\end{cases} 
\end{equation}
where $a_{uv}^k$ is equal to the number of times nodes $u$ and $v$ are graphlet-adjacent with respect to a given graphlet $G_k$.~\cite{graphletlaplacians}

\subsection{Novel graphlet-based network matrix representations}

In this study, we generalize the PPMI formula used in NLP, to perform network embeddings that consider local subgraphs. In particular, we use as the corpus all instances of a given graphlet in the network and then compute how frequently two nodes co-appear in these instances of a given graphlet in the network. Recall that a Graphlet Adjacency Matrix captures in how many instances of a given graphlet two nodes co-occur. Thus, to transform the raw co-occurrences to probabilities of co-occurrences, we apply to each Graphlet Adjacency Matrix the PPMI formula from NLP (see equation~\ref{eq:PPMI_NLP}). We call this generalization of the PPMI the \emph{Graphlet Pointwise Mutual Information (GPMI)}. To generate the Graphlet Adjacency Matrices, we use the implementation of Windels \textit{et al.} \cite{graphletlaplacians}

Formally, for two nodes, \textit{u} and \textit{v}, GPMI is defined as:
\begin{equation}
\label{eq:PPM_NLPI}
GPMI(u,v)= max \left(0, \log \frac{vol(A_k) \times A_k(u,v)}{D_k(u) \times D_k(v)}\right),
\end{equation}
where $vol(A_k)$ is the total number of instances of the given graphlet $k$ in the network, $A_k(u,v)$ is the number of instances of graphlet $k$ in which nodes $u$ and $v$ co-occur, and $D_k(u)$ and $D_k(v)$ are the numbers of instances of graphlet $k$ in which node $u$ and node $v$ occur, respectively.

In the DeepWalk closed-formula, the key parameter is the adjacency matrix of the network, A, in which the random walks of length \textit{T} are computed. Hence, to incorporate the graphlets into the DeepWalk closed formula, we propose to compute random walks in all  different Graphlet Adjacency Matrices corresponding to the nine graphlets with up to 4 nodes (illustrated in Figure~\ref{fig:graphlets}). In the Graphlet Adjacency Matrix,  $A_k$, the entries represent the numbers of times two nodes simultaneously participate in the given graphlet, $G_k$. This number can be very big, especially for higher-order (larger) graphlets. To address this, we introduce the binarized Graphlet Adjacency Matrix, $\tilde{A_k}$, by setting all the non-zero values to one. Then, we use as input for the DeepWalk closed-formula any of the nine binarized Graphlet Adjacency Matrices. We name this generalization of the DeepWalk closed formula the \emph{DeepGraphlet} for graphlet $G_k$: 

\begin{equation}
\label{eq:Deepgraphlets}
DeepGraphlet_k = max \left(0, \log(vol(\tilde{A_k})\left(\frac{1}{T}\sum_{r=1}^{T}(\tilde{D_k}^{-1}\tilde{A_k})^{r}\right)\tilde{D_k}^{-1})\right),
\end{equation}
where $vol(\tilde{A_k})$ is the volume of the binarized Graphlet Adjacency Matrix and is computed as $vol(\tilde{A_k})= \sum_{i} \sum_{j} \tilde{A_k}(i,j) $, $\tilde{A_k}$ is the binarized Graphlet Adjacency Matrix of the network, $\tilde{D}_k$ is the diagonal matrix of degrees of the given binarized Graphlet Adjacency Matrix and \textit{T=10} is the length of the random walks.

In both of our newly introduced network matrix representations, GPMI and DeepGraphlet, we have nine different graphlet-based representations of a network that correspond to the different connectivity patterns captured by each of the nine 2-node to 4-node graphlets.

\subsection{Homophily measures}

To assess if the unweighted matrix representations (Graphlet Adjacency Matrices and binarized Graphlet Adjacency Matrices) of a network $G=(V,E)$, $A_k$ and $\tilde{A_k}$, are homophilic (i.e., if the adjacent nodes share the same label), we use the two standard metrics: the edge homophily index and node homophily index.\cite{zhu2020beyond} The edge homophily index, $H_{edge}$, quantifies the proportion of edges connecting two nodes from the same class  and is defined as:
$$H_{edge}=\frac{ | (u,v) \in E: label(u)=label(v)|}{|E|}, $$ where $(u,v)$ denotes a pair of nodes, i.e. an edge, of the graph.
The node homophily index, $H_{node}$, quantifies the average proportion of the adjacent nodes that share the same label (class).\cite{zheng2022graph} Formally, it is defined as:
$$H_{node}=\frac{1}{|V|} \sum_{u \in V} \frac{|v \in N(u): label(u)=label(v)|}{|N(u)|},$$ where $N(u)$ is the set of the neighbors of node u.

We extended these indices for the weighted matrix representations of a network: PPMI, DeepWalk and their graphlet-based extensions. We do this to assess if nodes with stronger connections (i.e., higher weights) share the same label. If two nodes share the same label, we use the corresponding edge weight instead of 1. The weighted edge homophily index is defined as:

$$
H_{\text{edge}}^{\text{weighted}} = \frac{| \sum_{(u,v) \in E} w_{uv}: \text{label}(u) = \text{label}(v)|}{\sum_{(u,v) \in E} w_{uv}},$$ where $(u,v)$ is an edge between nodes $u$ and $v$ in the graph and $w_{uv}$ is the corresponding edge weight.
The weighted node homophily index is defined as:
$$
H_{\text{node}}^{\text{weighted}} = \frac{1}{|V|}  \sum_{u \in V} \frac{| \sum_{v \in N(u)} w_{uv}: \text{label}(u) = \text{label}(v)|}{\sum_{v \in N(u)} w_{uv}},$$ where $N(u)$ is the set of the neighbors of node u.


Note that since the original edge and node homophily indices are only applicable for unweighted matrix representations of a network (e.g., unweighted adjacency matrix), we also compute the Geometric Separability Index (GSI) \cite{thornton1998separability}, a less stringent version of the node homophily index, that can also be applied to weighted matrix representations of networks (e.g., GDV PPMI and DeepGraphlets). GSI is defined as the proportion of network nodes whose labels are the same as those of their first-nearest neighbors. If the network is unweighted, the GSI first computes the pairwise Euclidean distances of the network nodes to identify the first-nearest neighbor of each node.

We extend GSI, node and edge homophily indices (weighted and unweighted) for multi-label (biological) networks, in which a node may have more than one annotation. To do so, for each node, we compute the percentage of shared annotations with its neighbors.

\subsection{Non-negative matrix tri-factorization based embeddings}

Following our published work\cite{xenos2021linear}, we use an Orthonormal NMTF (ONMTF) framework to decompose the different representation of the networks (Adjacency, PPMI, DeepWalk and their graphlet-based extensions) and generate the corresponding network embedding spaces. In particular, given an input network matrix representation, \textit{X}, our ONMTF framework decomposes it into three non-negative matrix factors, \textit{E}, \textit{S} and $P^{T},$ as $X\approx ESP^{T}$, where $E \cdot S$ contains the vector representations of the entities of \textit{X} in the embedding space spanned by $P^{T}$, \textit{S} is a compressed representation of network \textit{X} and $P^{T}$ is the orthonormal basis of the embedding space. Importantly, the orthonormality constraint ($P^{T}P=I$) leads to independent, nonambiguous directions in the embedding space and improves the clustering interpretation of NMTF.\cite{ding2006orthogonal}
The decomposition is done by solving the following:
\begin{equation}
min_{G,S,P \geq0 } \lVert X- ESP^{T}\rVert_{F}^{2},P^{T}P=I
\end{equation}
where F denotes the Frobenius Norm.
This optimization problem is NP-hard \cite{ding2006orthogonal}, thus to solve it we use a fixed point method that starts from an initial solution and iteratively uses the multiplicative update rules \cite{ding2006orthogonal}, derived from the Karush-Kuhn-Tucker (KKT) conditions, to converge towards a locally optimal solution (see \cite{xenos2021linear} for details). To generate the initial $E$, $S$ and $P$ matrices, we use the Singular Value Decomposition (SVD) based strategy.\cite{Qiao2015} This strategy makes the solver deterministic and also reduces the number of iterations that are needed to achieve convergence (see \cite{Qiao2015} for more details). We stop the iterative solver after 500 iterations.  







\subsection{Linear separability of the embedding space}

An embedding space is considered to be linearly separable if the embedding vectors of nodes from different classes can be separated by hyperplanes. If the nodes belonging to different classes in the embedding space are linearly separable, then a linear classifier can classify them into their respective classes as accurately as a non-linear classifier. Hence, to assess the linearity of an embedding space, we perform node classification on single-label and multi-label networks by using Support Vector Machines (SVM) with both linear (Euclidean) and non-linear (Radial Basis Function, RBF) kernel. In addition, we compare their performance with that of the Random Forest (RF) classifier, a state-of-the-art non-linear method.\cite{breiman2001random} To evaluate the accuracy of the classifiers, we use 10-fold cross-validation and compute the corresponding weighted F1-score for each classifier. 

We say that an embedding space is \textit{sufficiently linearly separable}, if the node classification F1-scores of the linear SVM (Euclidean kernel) are on par, or surpass those of the non-linear methods (SVM RBF and RF). To demonstrate that there are no statistically significant differences in the performances of the linear and non-linear classifiers, we compare the distributions of their F1 scores by using the Mann-Whitney U test. If in addition to having no statistically significant differences, these F1-scores are also greater than 0.8, we say that the embedding space is \textit{fully linearly separable}. Finally, if the F1-scores of the non-linear classifiers are higher than those of the linear methods, we say that the embedding space is \textit{non-linearly separable}.

\subsection{Random partition graph model}

To demonstrate the relationship between the homophily level of the input network matrix representation and the linear separability in the embedding space under ideal conditions, without noisy data, we simulate data using the random partition graph model.\cite{fortunato2010community} The main parameters of this model are the number of partitions (representing communities), the size of each community, the probability of connections between nodes within the same community ($p_{in}$, intra-edges), and the probability of connections between nodes from different communities ($p_{out}$, inter-edges). We set the number of nodes to 1,000 distributed across 5 communities and we vary the $p_{in}$ probability from 0.05 to 1 in increments of 0.05 and the $p_{out}$ from 0 to 1 in increments of 0.05, resulting in the total of 420 networks. Note that starting both probabilities at 0 would result in networks with no edges, making it meaningless to apply the homophily measures. For each simulated network, we compute the homophily level of the input network matrix representation: adjacency matrix, DeepWalk closed matrix formula and LINE closed matrix formula. Then, we use our NMTF-based framework to generate the embedding space, and for each embedding space, we perform node classification by using SVM with the Euclidean kernel. Finally, we compute the Pearson Correlation Coefficient between the homophily levels of the input network matrix representations and the F1-node classification scores.


\subsection{Downstream analysis tasks}

To demonstrate that higher linear separability of classes in the embedding space leads to better downstream analysis results, we perform two different experiments: one for single-labeled networks and another for multi-labeled biological networks. For the single-labeled networks, we predict the label of the nodes based on their distances in the embedding space. For the biological networks, where nodes (genes) can have multiple annotations or remain unannotated due to unknown functions, a common downstream analysis task involves uncovering functional network modules — sets of genes that collectively perform higher-level biological functions.

\paragraph{\indent Label prediction in single label networks:} 

In a single-labeled network, each node is assigned one label, and two nodes that have the same label belong to the same class. A common downstream task in an embedding space is to classify a node based on its cosine similarity from nodes with known labels. The cosine similarity is a measure of similarity between two vectors that considers the angle between them rather than their length. Here, we evaluate if the embedding spaces of single-labeled networks are organized coherently with respect to their node classes. To do so, we formulate the problem as a multi-class classification problem, in which we aim to classify two nodes with the same label based on their cosine similarity. The ability of the cosine similarity between the embedding vectors of nodes to correctly group nodes of the same class is evaluated by using the ROC curve (AUROC) analysis.\cite{bradley1997use} Since we have more than one node class (label), we use the one-vs-rest strategy to generalize the binary classification task to the multi-class problem. In this strategy, we use one binary classifier for each possible class and compute its corresponding AUROC score. Then, to account for the imbalances between the different classes, we calculate the weighted AUROC score.



\paragraph{\indent Functional module discovery in biological networks:}

In biological networks, a common downstream analysis task is uncovering functional network modules, i.e., sets of genes that together perform higher-level biological functions. After embedding a biological network, this is typically done by clustering the genes based on the proximity of their embedding vectors (Euclidean distance) in the embedding space. Subsequently, the biological relevance of the obtained gene clusters is measured by assessing if the genes that cluster together share biological annotations statistically significantly more often than expected at random; this is done by over-representation (enrichment) analysis, detailed below.

To assess whether genes that are close in the embedding space of the PPI and COEX networks perform higher-level biological functions, we cluster the genes by using \textit{k}-means clustering on the node (gene) embedding vectors obtained from the orthogonal space, $E \cdot S$, of the ONMTF. The number of clusters, \textit{k}, is determined by using the heuristic rule of thumb, $k = \sqrt{\frac{n}{2}}$, where \textit{n} is the number of data points, i.e. nodes in the network.\cite{kodinariya2013review} We evaluate the biological relevance of the obtained gene clusters by measuring the percentages of clusters that are enriched in Reactome Pathways (RP) terms. The probability that an annotation is enriched in a cluster is computed by using the hypergeometric test (sampling without replacement strategy):



\begin{equation}
p=1-\sum_{i=0}^{X-i} \binom{K}{i}  \binom{M-K}{N-i} / \binom{M}{N},
\end{equation}
where \textit{N} is the number of annotated genes in the cluster, \textit{X} is the number of genes in the cluster that are annotated with the given annotation, \textit{M} is the number of annotated genes in the network and \textit{K} is the number of genes in the network that are annotated with the given annotation. An annotation is considered to be statistically significantly enriched if its enrichment \textit{p}-value, after Benjamini and Hochberg correction \cite{benjamini1995controlling} for multiple hypothesis testing, is lower than or equal to $5\%$.

We measure the functional coherence of the clustering by computing the percentage of genes that are grouped together and share functions, i.e., that have at least one of their annotations enriched in their clusters, over all annotated genes --- we term it ``gene coverage''. In addition, we compute if the space captures all the possible functions of the genes, i.e., the percentage of annotations that are enriched in at least one of the gene clusters --- we term it ``functional coverage''.  Note that we report the results over the set of genes with known annotations. For the number of annotated nodes in each of these networks, see Supplementary Tables \ref{stable:bio_annotated_nodes}, \ref{stable:kegg_annotated_nodes} and \ref{stable:rp_annotated_nodes}.

\section{Results}

To generate embeddings that simultaneously capture node neighborhood similarity and node topological similarity, we use the nine 2- to 4- node graphlets (Fig. S1) as follows. To randomly walk between similarly wired nodes in the network, from a given node we can visit any other node that simultaneously participates in the selected graphlet (specified by us) that the given node participates in;  we do this for all $x$ graphlets that the given node participates in to generate $x$ different matrices and we repeat this for each node of the network. That is, the connectivity information is captured by the family of Graphlet Adjacency matrices ($G_{adj}$) \cite{graphletlaplacians}, where $G_0$ is the traditional adjacency matrix (detailed in Methods). We extend the DeepWalk and the Line matrix closed formulae to account for these new graphlet based random walks captured by the Graphlet Adjacency matrices, hence defining the new \textit{DeepGraphlets} family and the new \textit{Graphlet Pointwise Mutual Information (GPMI)} family of closed formulae, respectively, defined for each of the 9 graphlets (detailed in Methods). For a given graphlet, $G_k$, we term the DeepWalk graphlet-based extension as $DeepGraphlet_{G_k}$ and the LINE graphlet-based extension as $GPMI_{G_k}$ (for details, see ``Graphlet-based network embeddings'' in Methods). Note that for all of the new matrix representations, those based on graphlet $G_0$ correspond to the original methods, with which we compare our new methods; the rest of our new graphlet-based extensions (termed ``higher-order graphlets") capture both the neighborhood-based similarity (as in the competing methods) and the topological similarity (beyond only the direct neighborhood). In addition, we compare our new graphlet-based extensions with the baseline Graphlet Adjacency matrices and with two network representations that rely only on the topological similarity between the nodes, the Graphlet Degree Vector (GDV) similarity matrix and the GDV PPMI matrix \cite{xenos2021linear} (for details, see ``Graphlets and Graphlet Adjacency'' in Methods).


We apply our new graphlet-based network representations on networks in several application domains.  In particular, we apply them on systems-level molecular interaction networks, in which nodes corresponds to genes (or equivalently, to proteins, as gene products), which can have multiple annotations (labels).  We use the protein-protein interaction (PPI) and gene co-expression (COEX) networks of three species, \textit{Homo sapiens}, \textit{Saccharomyces cerevisiae}, and \textit{Schizosaccharomyces pombe}, resulting in six molecular networks. The sizes of these six molecular networks are presented in Table \ref{table:bio_netstats}. In addition, we also apply our methodology on seven commonly used single-label networks from different domains: the USA air-traffic network \cite{ribeiro2017struc2vec}, the Coauthor computer science (CS) network \cite{shchur2018pitfalls}, the CORA and the CiteSeer \cite{sen2008collective} citation networks, the Wiki-CS hyperlinks network \cite{mernyei2020wiki} and the two Wikipedia page–page networks (Chameleon and Squirrel).\cite{multi-scale} Note that two of the single-label networks, Chameleon and Squirrel, are heterophilic based on their standard adjacency matrix representation (for details, see ``Single-label networks'' in Datasets in Methods). The statistics of the these networks are presented in Table \ref{table:netstats}. 


\subsection{Graphlet random walk-based matrices yield more homophilic network representations}

To demonstrate that our new graphlet-based extensions of LINE and DeepWalk yield more homophilic representations of the networks than the original baseline methods (that also correspond to our $G_0$ extension, as detailed above), we measure the level of homophily of the different matrix representations of six molecular multi-label networks and seven single-label networks (detailed above). Traditionally, in unweighted networks represented by their standard adjacency matrices, homophily is measured by the edge homophily index (the proportion of edges connecting two nodes from the same class) and the node homophily index (the proportion of adjacent nodes from the same class).\cite{zheng2022graph} In weighted networks, it is  measured by the Geometric Separability Index (GSI) \cite{thornton1998separability}, a simplified measure that compares the label of a node only with the label of its closest neighbor. In this study, we extend these measures to quantify the homophily/heterophily in any matrix representation of the network (weighted or unweighted). In the case of the molecular networks, we annotate the genes with level 1 Gene Ontology Biological Process (GO BP) terms \cite{ashburner2000}, which represent higher-level biological functions. Since the functions (annotations) of all genes are not known, we report the results over the set of genes with known annotations. For the number of annotated nodes in each of these networks, see Supplementary Table \ref{stable:bio_annotated_nodes}.

\begin{figure}[H]
	\begin{center}
		\begin{tabular}{c c}
		\hspace{-6.5cm}\textbf{(A)} & \hspace{-6.5cm}\textbf{(B)} \\
		\includegraphics[width=8cm]{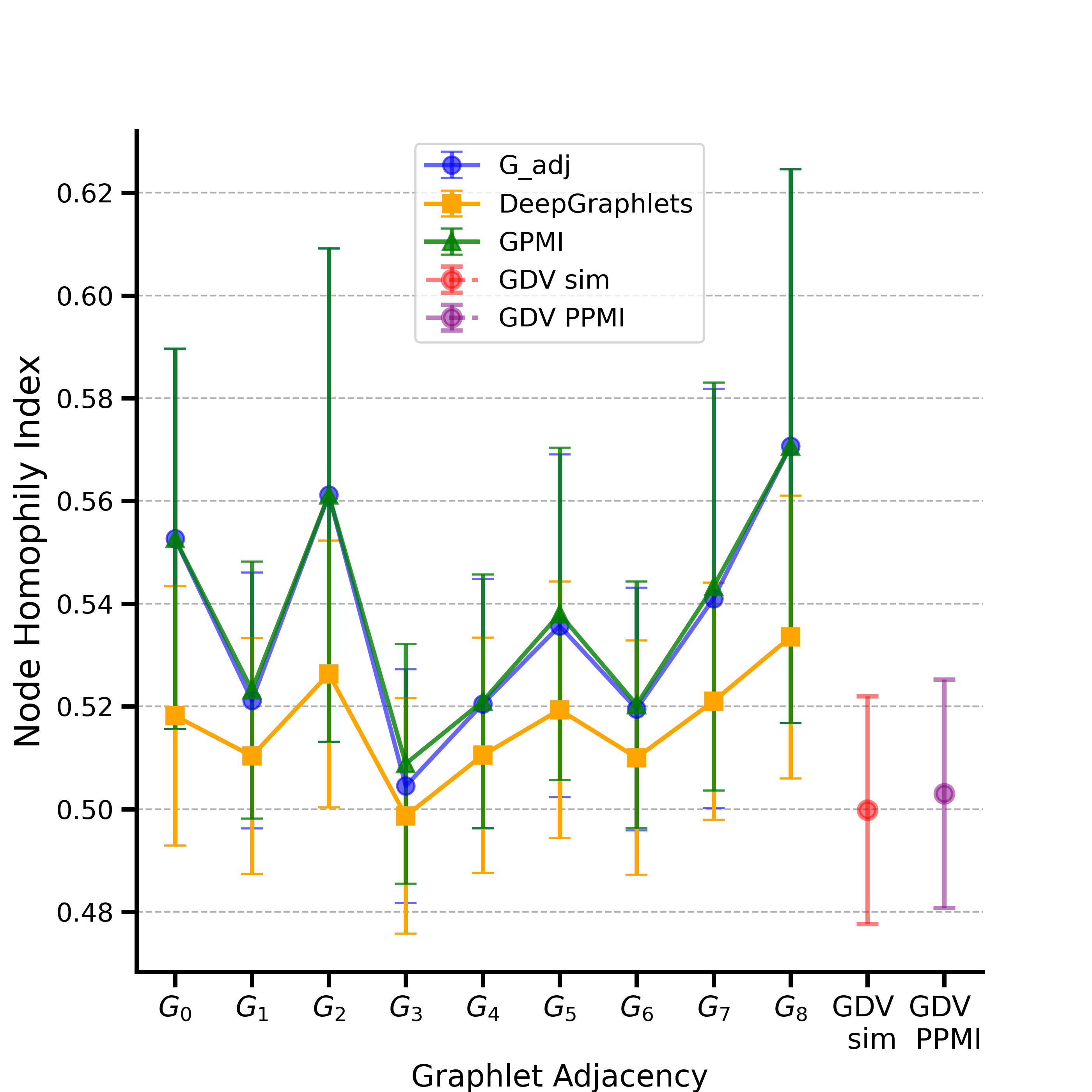} &
		\includegraphics[width=8cm]{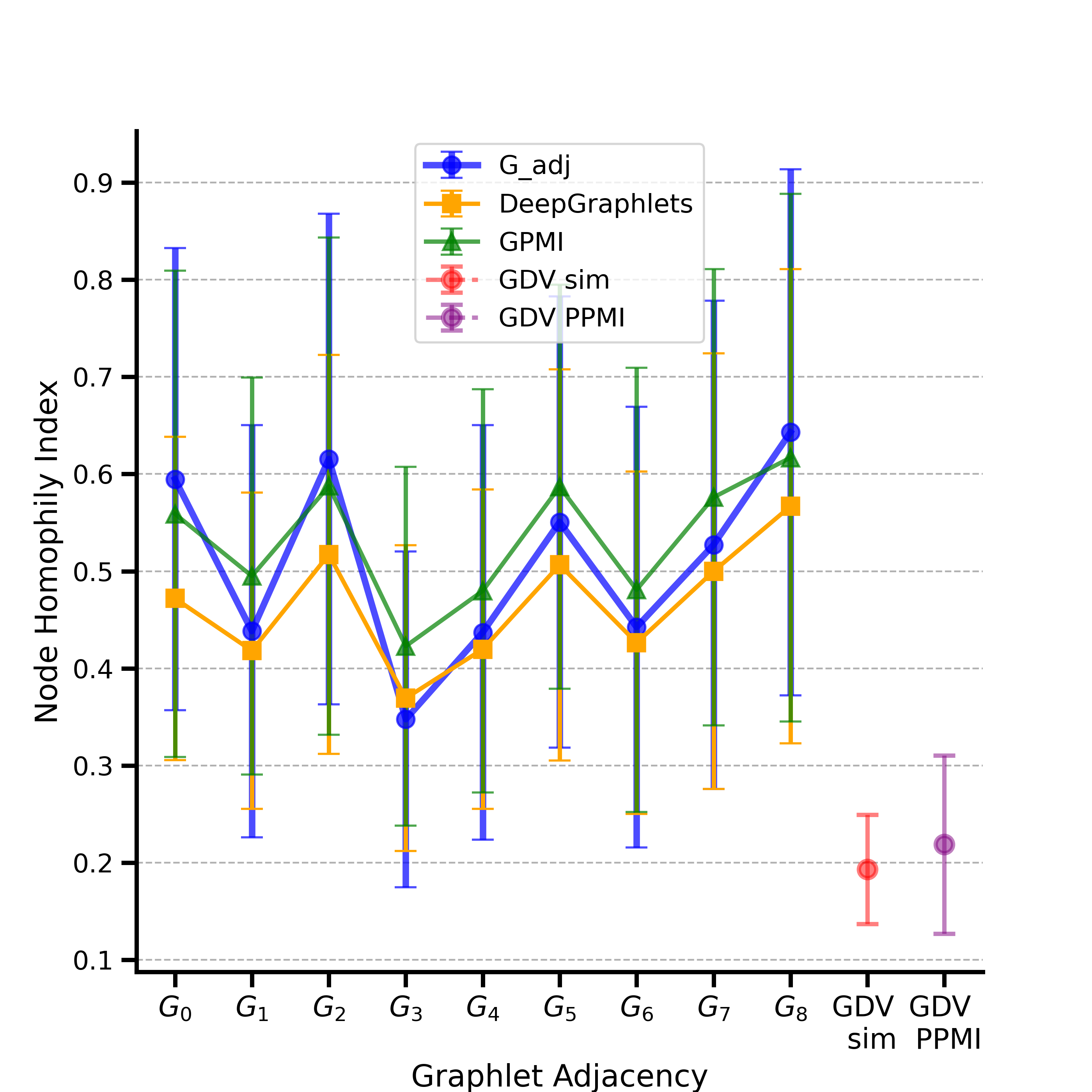} \\
        \hspace{-6.5cm}\textbf{(C)} & \hspace{-6.5cm}\textbf{(D)} \\
        \includegraphics[width=8cm]{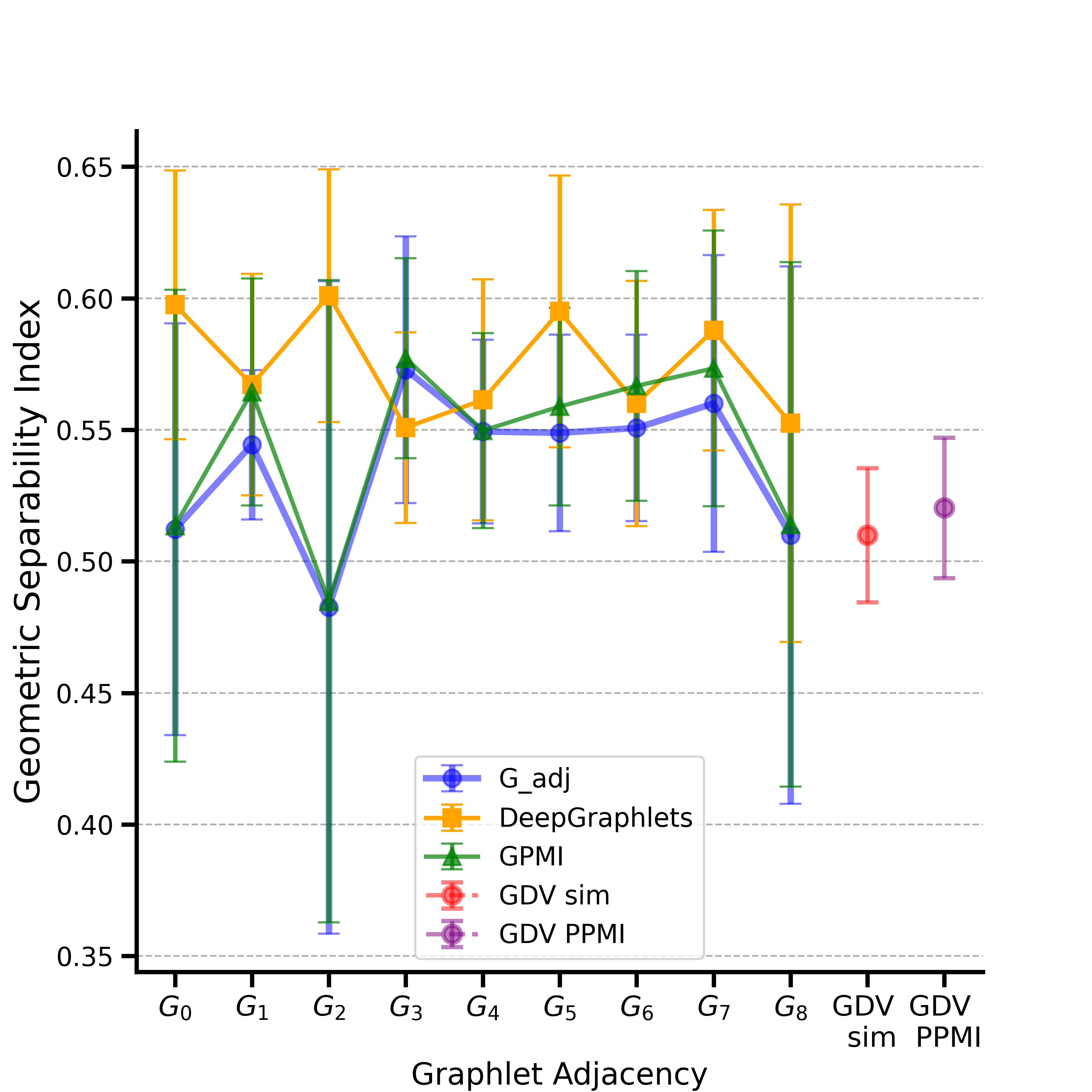} &
        \includegraphics[width=8cm]{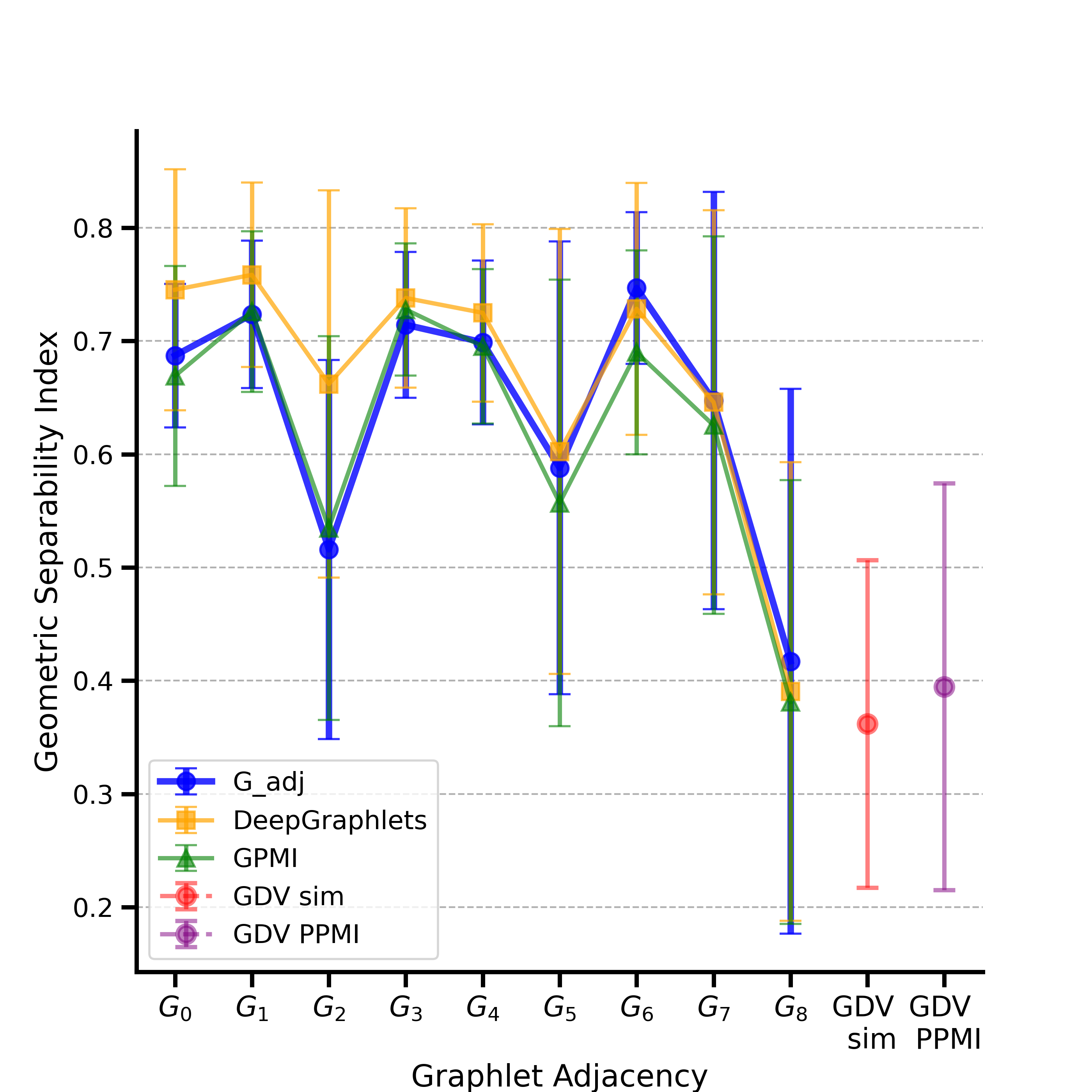} \\
		\end{tabular}
	\end{center}
    \caption{{\bf Graphlet-based network matrix representations lead to more hopophilic representations.} In Panel {\bf A}, for each graphlet (x-axis) and method (color-coded), the line plot shows the average, over the six biological multi-labeled networks, node homophily index and the standard deviation. Panel {\bf B} shows the same, but on average over the seven single-labeled networks. In panel {\bf C}, for each graphlet (x-axis) and method (color-coded) the line plot shows the average Geometric Separability Index (y-axis), over the six biological multi-labeled networks, along with the standard deviation. Panel {\bf D} shows  the same, but on average over the seven single-labeled networks. \label{fig:graphlets_representations}}
\end{figure}

We observe that for both the multi-label and the single-label networks, there is at least one higher-order graphlet-based network representation that is more homophilic (i.e., having greater GSI, node homophily and edge homophily indices) than the baseline adjacency matrix, DeepWalk and LINE matrix closed formulae (see Figure \ref{fig:graphlets_representations}). In addition, over all  networks and over all graphlets, our new $DeepGraphlet_{G_k}$ and $GPMI_{G_k}$ representations that capture both topological and neighborhood-based similarity yield more homophilic representations than GDV similarity matrix and GDV PPMI matrix that are based solely on topological similarity (see Figure \ref{fig:graphlets_representations}). Among the different network representations, graphlet $G_2$ and graphlet $G_8$ extensions of GPMI and $G_{Adj}$ lead to the most homophilic representations in terms of the node and edge homophily index (see Panels A and B of Figure \ref{fig:graphlets_representations}). In addition, our new DeepGraphlets yield the most homophilic representations in terms of the GSI (see Panel C and D of Figure \ref{fig:graphlets_representations}): $DeepGraphlet_{G_1}$ in case of single-label networks, and $DeepGraphlet_{G_2}$ in case of multi-label networks.

In summary, for both the multi-label and the single-label networks, our new graphlet-based matrix representations of the networks are more homophilic than the original representations that capture either only the direct neighbourhood similarity, or only the topological similarity. In the following sections, we assess if embedding the networks by factorising these, more homophilic matrix representations of networks leads to more linearly separable network embedding spaces.


\subsection{Graphlet random walk-based network representations lead to linearly separable embedding spaces}


We examine if our graphlet-based matrix extensions detailed above can be applied to obtain the state-of-the-art network embedding methods yielding more linearly separable network embedding spaces. That is, we examine if they can be applied to result in the embedding vectors of nodes from different classes that can be separated by hyperplanes. To do that, first we generate the network embedding spaces by factorizing all the different matrix representations (DeepGraphlet, GPMI and Graphlet Adjacency matrices) of the seven single-label networks detailed above.  Recall that this involves 9 graphlet-based matrix representations (because there are 9 up to 4-node graphlets) per method (for the 3 methods, DeepGraphlet, GPMI and Graphlet Adjacency matrices) for each of the 7 single-label networks, hence yielding the total of $9 \times 3 \times 7 = 189$ embedding spaces. We do the matrix factorizations by utilizing the ONMTF framework (detailed in ``Non-negative matrix tri-factorization embeddings'' in Methods). Similarly, we generate the embeddings of the six molecular interaction networks detailed above (i.e., 9 graphlet-based representations per method per network, yielding the total of $9 \times 3 \times 6 = 162$ embedding spaces). Recall that if the classes in an embedding space are linearly separable, then a linear classifier should classify the nodes into their respective classes as accurately as a non-linear classifier. Hence, we assess the linearity of an embedding space produced by our methods by comparing the resulting node classification weighted F1-scores to those obtained by Support Vector Machines (SVM) with linear kernel (Euclidean) and non-linear kernel (Radial Basis Function, RBF). In addition, we assess it by comparing with that of the Random Forest (RF) classifier, a state-of-the-art non-linear method \cite{breiman2001random} (for details, see ``Linear organization of the embedding space'' in Methods). We term a space as ``\textit{sufficiently linearly separable}'' if the node classification weighted F1-scores of the linear classifiers are on par, or better than those of the non-linear classifiers. In addition, if these weighted F1-scores are also greater than 0.8, we term the space as being ``\textit{fully linearly separable}''. Finally, if the weighted F1-scores of the non-linear classifiers are larger than those of the linear methods, we term the space as being ``\textit{non-linear}''.

As shown in Table \ref{table:all_networks_node_classification},
in 9 out of the 13 networks, the corresponding embedding spaces are sufficiently linearly separable, since the node classification F1-scores of the linear SVM outperform those of the non-linear SVM (RBF kernel). In 3 of these networks (Cora, Wikipedia CS and CS Co-author), the corresponding embedding spaces are fully linearly separable, since the F1-scores exceed 0.8 (average weighted F1-score of 0.85, see left panel in Supplementary Figure \ref{sfig:node_classification}) and there are no statistically significant differences between the classification results of the linear and the non-linear approaches (Mann Whitney U-test between their distributions of F1 scores having $p-values >= 0.05$). Only in four single-label networks (USA air-traffic, Chameleon, Squirrel and CiteSeer) the embedding spaces are not linearly organized, since the RF classifier outperforms the linear and the non-linear SVM by at least 14 \% in the node classification F1-scores (see the right panel of Supplementary Figure \ref{sfig:node_classification}).  


Regardless of the classifier, the best results over all networks are obtained with DeepGraphlets with an average F1-score of 0.66, followed by GPMI (0.647 average F1-score) and then by the raw $G_{Adj}$ (0.623 average F1-score). Hence, our new DeepGraphlets and GPMI representations that employ random-walks to diffuse the information on the input adjacency matrix lead to more separable embedding spaces than the $G_{Adj}$ representations that embed directly the adjacency matrices. For the three networks with the embedding spaces being fully linearly separable, the highest F1-scores are obtained with the standard adjacency matrix representation, $G_0$ (see Supplementary Table \ref{stable:all_networks_node_classification_index}). This means that the node classes are already linearly separable in the embeddings generated by the standard adjacency matrix and hence, the higher-order graphlets are not needed to disentangle them. On the other hand, in three out of the six networks with the embedding spaces sufficiently linearly separable, i.e. for \emph{Pombe} COEX, \emph{Cerevisiae} PPI and \emph{Homo sapiens} PPI network, the highest F1-scores are achieved with the embeddings generated by utilizing the higher-order graphlets in our extended methods, $GPMI_{G_3}, DeepGraphlets_{G_8}$ and $DeepGraphlets_{G_2}$, respectively (see Supplementary Table \ref{stable:all_networks_node_classification_index}). In the four networks yielding embedding spaces that are non-linearly separable, that include the two standard heterophilic networks, our graphlet-based extensions of Deepwalk ($DeepGraphlet_{G_2}$, $DeepGraphlet_{G_4}$ and $DeepGraphlet_{G_6}$) yield the maximum weighted F1-score of $0.72$, outperforming the original DeepWalk method by almost 8 \%. Hence, in the networks with the resulting embedding spaces that are sufficiently linearly separable or non-linear (the RF classifier outperforms the linear classifiers), our higher-order DeepGraphlets and GPMI lead to embeddings that better  separate the different node classes.

In conclusion, we demonstrate in single-label and multi-label networks that our proposed graphlet-based embeddings lead to better class separability in the embedding space compared to the methods that rely only on topological similarity, or only on the direct neighbourhood similarity. However, despite these improvements, some embedding spaces remain non-linear. Hence, in the following section, we investigate if there is an intrinsic property in the input network matrix representation that is related with the linear separability of the resulting embedding space.


\subsection{More homophilic network representations lead to linearly separable embedding spaces}

We hypothesize that the more homophilic the input network matrix representation, the more linearly separable are the node classes in the resulting embedding space. To demonstrate this under ideal conditions, without noisy data, we generate synthetic data from the random partition graph model.\cite{fortunato2010community} This model allows the generation of random graphs that exhibit clustering and modular organization akin to real-world networks. We simulate 420 networks, each containing 1,000 nodes distributed across five communities (for details, see ``Random partition graph model'' in Methods). We represent each network with the DeepWalk closed matrix formula, the LINE closed matrix formula and the standard Adjacency matrix. For each network's matrix representation, we compute its node homophily index, edge homophily index and GSI. Subsequently, we generate the network embedding spaces by factorizing their aforementioned matrix representation by using the NMTF framework (see ``Non-negative matrix tri-factorization embeddings'' in Methods) and for each embedding space, we compute the node classification F1-score with the linear SVM (Euclidean kernel). We observe a positive correlation between the homophily level of the input matrix representation and the node classification F1-scores: 0.68 Pearson Correlation Coefficient (PCC) for GSI and 0.19 PCC for node and edge homophily indices, all statistically significant (see Table \ref{table:random_partition_model}). To assess if this observation holds for our real networks, we computed the correlations between the homophily indices of the network matrix representations of our 13 real networks and the node classification F1-scores in the corresponding embedding spaces. In the single-label networks, we observe 0.44 PCC for GSI and approximately 0.36 PCC for node and edge homophily indices, all statistically significant (see Table \ref{table:random_partition_model}). In the multi-label molecular networks, we observe 0.46 PCC for GSI, 0.53 PCC for the node homophily index and  0.44 PCC for the edge homophily index, all statistically significant (see Table \ref{table:random_partition_model}).

Hence, we verify our hypothesis in simulated and real world networks that the more homophilic the input network matrix representation, the more linearly separable are the node classes in the resulting embedding space. In the following section, we show that our new graphlet-based matrix representations of networks, which are more homophilic than their traditional matrix representations, also yield network embeddings resulting in better results in downstream analysis tasks.


\subsection{Graphlet-based embeddings lead to better results in downstream analysis tasks}

\begin{figure}[H]
	\begin{center}
		\begin{tabular}{c c}
		\hspace{-6.5cm}\textbf{(A)} & \hspace{-6.5cm}\textbf{(B)} \\
        \includegraphics[width=8cm]{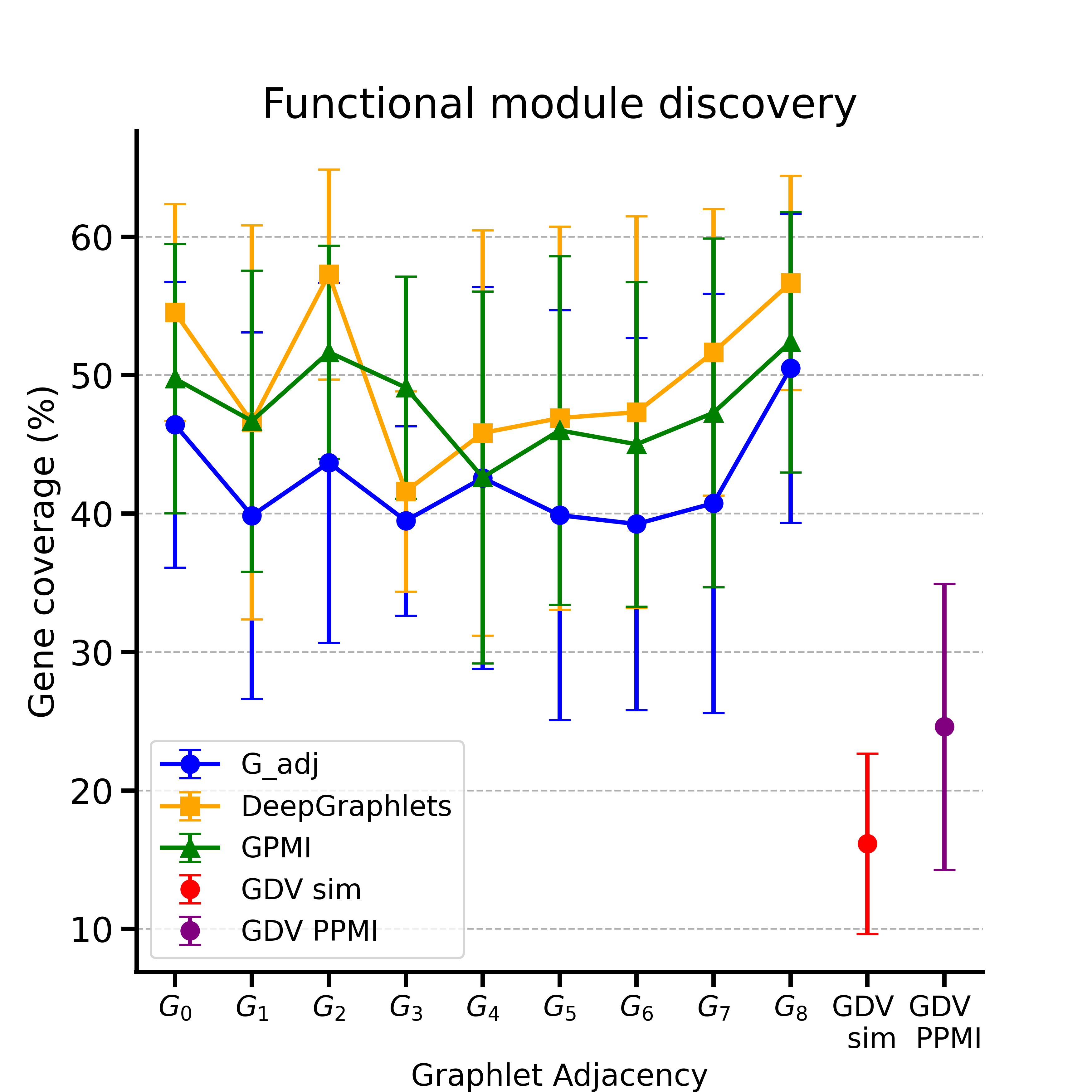} &
        \includegraphics[width=8cm]{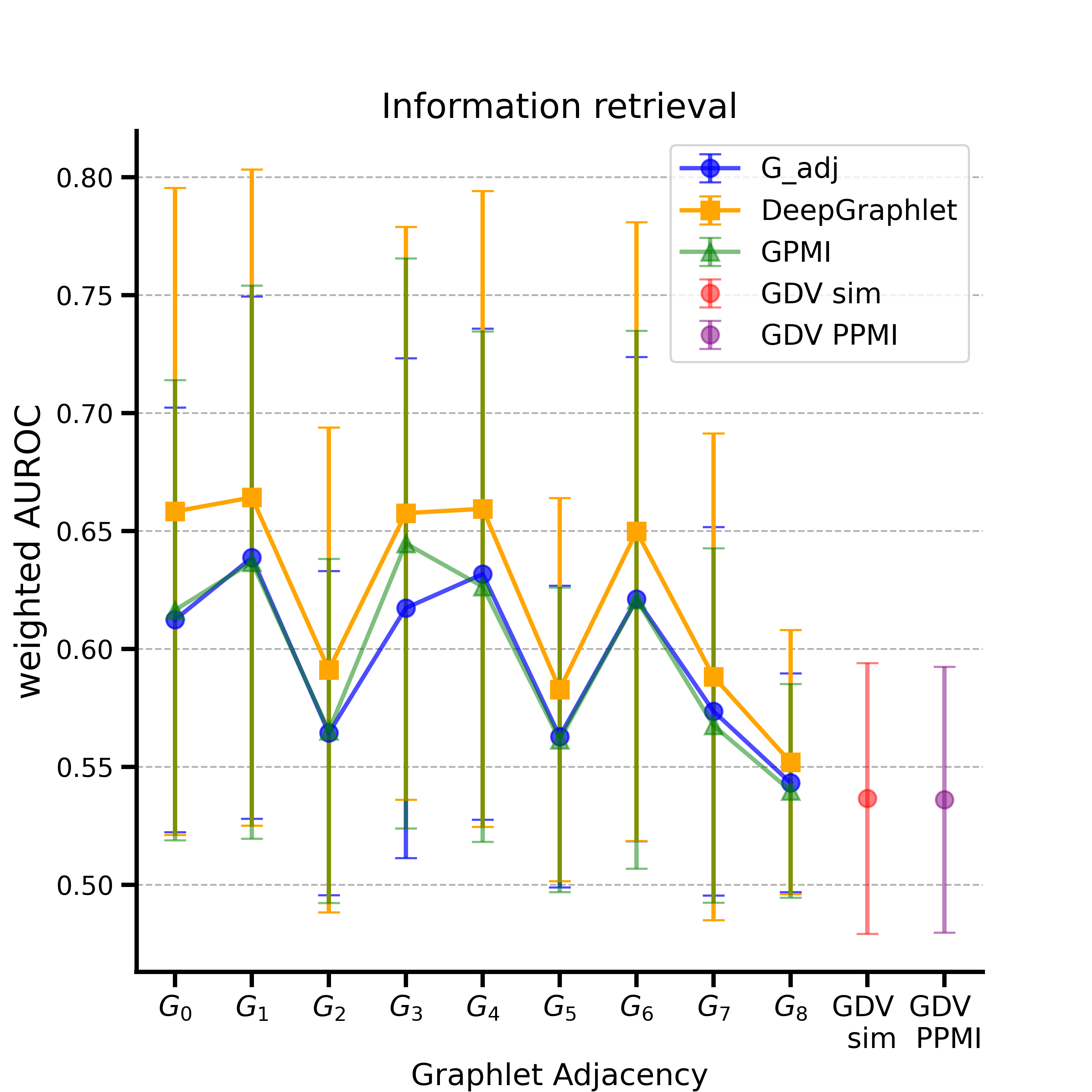} \\
		 \\
		\end{tabular}
	\end{center}
    \caption{{\bf Graphlet-based embeddings lead to better results in downstream analysis tasks.} Panel {\bf A} presents the results of the functional module discovery in gene embedding spaces of the biological multi-labeled networks. In particular, for each graphlet (x-axis) and for each method (color-coded) the line plot shows the average percentage, over the six molecular networks, of annotated genes in the clusters that have at least one Reactome Pathway (RP) term enriched in their clusters, along with the standard deviation (y-axis). Panel {\bf B} presents the results of the label prediction based on the cosine similarity in the embedding space of the single-labeled networks. For each graphlet (x-axis) and for each method (color-coded), the line plot shows the average weighted AUROC score (y-axis) over the seven single-labeled networks and the standard deviation. \label{fig:downstream_analysis_tasks}}
\end{figure}

We assess the performance of our new graphlet-based network embeddings in downstream analysis tasks for single-label and multi-label real-world networks. In single-label networks, a common information retrieval task is to predict the label of a node based on the label of its most semantically similar node, defined by the largest cosine similarity of their embedding vectors. Therefore, we evaluate whether our new graphlet-based network embedding spaces improve information retrieval by using cosine similarity of the node embedding vectors in the same way. To do so, we assess if nodes whose embedding vectors have high cosine similarity have the same label. Formally, for each network, we compute its weighted area under the ROC curve (AUROC) \cite{bradley1997use} between the ground truth (label) and the prediction score (cosine distance) over all the pairs of nodes in the network. For all the matrix representation families (DeepGraphlets, GPMI and $G_{Adj}$), our extensions based on higher-order graphlets $G_1$, $G_3$, $G_4$ and $G_6$ lead to embedding spaces that allows better information retrieval based on the cosine similarities of the nodes than the original matrix representations that are based on the standard adjacency matrix, $G_0$ (see Panel B in Figure \ref{fig:downstream_analysis_tasks}). In particular, our graphlet-based extensions of DeepWalk yield the most suitable embedding spaces for this downstream analysis task, with $DeepGraphlet_{G_1}$ achieving the highest weighted-AUROC (0.669 on average over the seven networks), followed by $DeepGraphlet_{G_3}$ (with weighted-AUROC of 0.663) and then by $DeepGraphlet_{G_4}$ (with weighted-AUROC of 0.659).  Also, we observe that our newly introduced familly of methods (DeepGraphlets, GPMI) outperform the ones that rely only on the topological similarity, GDV similarity matrix and the GDV PPMI matrix, which have, on average, 0.33 weighted AUROC score. 


In addition, we assess the performance of our graphlet-based embeddings in downstream analysis tasks in molecular, multi-label networks, in which nodes (i.e., genes) may lack annotations, or have overlapping annotations. A standard task is uncovering functional network modules, i.e., sets of genes that together perform higher-level biological functions. After embedding a biological network, this is typically done by clustering the genes based on their proximity (Euclidean distance) in the embedding space. Then, we assess if genes that are clustered together significantly share biological annotations by using over-representation (also called ``enrichment") analysis, that accounts for the incompleteness, noisiness and overlapping of biological annotations. To perform this analysis, we annotate genes in the molecular networks with Reactome Pathway (RP) terms, GO BP terms and KEGG pathways and we report the percentage of enriched functions (\textit{functional coverage}), which quantifies the number of functions over-represented (enriched) in some parts of the embedding space, and the percentage of enriched genes (\textit{gene coverage}), which quantifies the number of genes coding for proteins with these enriched functions (for details, see ``Functional module discovery in biological networks'' in Materials and Methods). However, since in the next section we correlate the linearity of the embedding spaces with the performance in the downstream analysis tasks, we present here the results when using RP terms in the downstream analysis; we do this to minimize circularity of using GO BP terms both to measure the linearity of the space (for details, see the second section of the Results and discussion) and to assess the performance in the downstream analysis tasks. In the Supplementary Material, we also present the results of using GO BP terms and KEGG pathways in downstream analysis, which are consistent with the results obtained using RP, presented below.

We find that over all molecular networks, there is always a graphlet-based extension of LINE and DeepWalk that outperforms the original method (corresponding to the adjacency matrix of the network, $G_0$) in terms of functional and gene coverages for RP terms (see Panel A in Figure~\ref{fig:downstream_analysis_tasks} and Supplementary Figure~\ref{sfig:geneenri}). The best results are obtained by DeepGraphlets, followed by GPMI and then by the raw Graphlet Adjacencies. This further verifies that the embeddings based on the DeepGraphlets and GPMI that employ random-walks to diffuse the information on the input adjacency matrix capture more biological information than the $G_{Adj}$ family that embed directly the adjacency matrices.\cite{xenos2021linear} The highest gene and functional coverage occurs with our $DeepGraphlet_{G_2}$ and $DeepGraphlet_{G_8}$, the two dense graphlet-based extensions (i.e. the 3-node and 4-node cliques) of the DeepWalk closed formula, which we have already shown to yield the most homophilic network matrix representations of molecular networks. In particular, the average gene coverage for DeepWalk is $54.53 \%$, whereas for $DeepGraphlet_{G_2}$ and $DeepGraphlet_{G_8}$ the average gene coverages are $57.28 \%$ and $56.67 \%$, respectively. The better performance in functional modules discovery of these interconnected graphlets is in line with the biological hypothesis that functions are performed by densely interconnected genes. For instance, proteins associated with the same molecular functions often form protein complexes, which are represented as highly-connected modules (dense subgraphs) within the PPI network \cite{chen2006detecting}. Finally, over all molecular networks and all graphlets, our new network embedding approaches that leverage both topological and direct neighborhood-based similarity ($DeepGraphlet_{G_k}$ and $GPMI_{G_k}$) yield embedding spaces that uncover more biologically coherent functional network modules than the embedding spaces that are based only on topological similarity (GDV similarity matrix and GDV PPMI matrix); see Panel A in Figure~\ref{fig:downstream_analysis_tasks} and Supplementary Figure~\ref{sfig:geneenri}. The other annotations (GO BP and KP) show consistent results, as shown in Supplementary Figures \ref{sfig:Kegg_downstream_analysis_tasks} and \ref{sfig:go_bp_downstream_analysis_tasks}. 

In conclusion, we demonstrate that our new graphlet-based network embeddings, that are based on factorizing more homophilic network matrix representations, yield better results in two downstream analysis tasks, one in single-label networks and one multi-label networks. In the next section, we show the connection between these improved results and our more homophilic graphlet-based matrix representations of the networks.  

\subsection{More homophilic network representations lead to better results in downstream analysis tasks}

To demonstrate that the more homophilic the input network matrix representation, the better the results of the downstream analysis tasks, we compute the Pearson’s correlation coefficients (PCC) between the homophily measures and the information retrieval results (in single label networks) and the functional module discovery (in multi-label molecular networks) results. For the 7 single-labeled networks, as shown in Table \ref{table:downstream_analysis_correlations}, all the homophily measures are statistically significantly correlated with the information retrieval task, having the following PCCs: 0.17 PCC for GSI, 0.61 PCC for node homophily and 0.57 PCC for edge homophily. For the 6 molecular, multi-labeled networks, all the homophily measures are also statistically significantly correlated with the functional module discovery, having the following PCCs: 0.36 PCC for GSI,  0.49 PCC for node homophily and 0.36 PCC for edge homophily.

In conclusion, we demonstrate that: (i) our graphlet-based matrix representations of the networks are more homophilic than the traditional ones; (ii) the more homophilic the matrix representation of the network, the more linearly organized the resulting embedding space, and (iii) the better the downstream analysis results in the embedding space.

\section{Discussion}

We introduce novel network embedding methods built on graphlet-based random walk matrix representations of networks; importantly, they capture both higher-order topological and direct neighbourhood information. Our novel matrix representations, alongside the traditional ones that capture either only the higher-order topological or only the direct neighborhood-based node similarity (but not both), enable us to explore the relationship between the intrinsic properties of the given network matrix representation, the topology of the resulting embedding space and the downstream analysis results in the embedding space. We demonstrate, on synthetic and 13 real-world networks from several application domains (including six multi-labeled and seven single-labeled networks), that the more homophilic the network representation, the more linearly separable the corresponding network embedding space, yielding better downstream analysis results.

It is known that heterophilic networks present challenges for graph ML methods, including random-walk based network (node) embeddings and GNNs. Improving the performance of downstream analysis tasks in heterophilic data is an active area of research \cite{morris2019weisfeiler,cotta2021reconstruction,bouritsas2022improving}, yielding a plethora of increasingly complex architectures. In this study, we propose shifting the focus from developing more complex ML models to elucidating the underlying properties of the data and exploiting these properties to construct explainable and environmentally sustainable (i.e., computationally efficient) models. To that end, we demonstrate that our newly introduced graphlet-based representations enable us to embed networks into more linearly separable spaces, consequently allowing for better mining of the networks by simplifying otherwise computationally-intensive machine learning models. Notably, in 9 out of the 13 studied networks, by using simple linear methods, we can achieve node classification F1-scores comparable to those of complex non-linear machine learning methods. For the four networks where the resulting embedding spaces are non-linear (the non-linear classifier outperform the linear classifier), our higher-order DeepGraphlets yield embeddings that better disentangle the different classes leading to $\approx 8\%$ improvement in the node classification F1-scores. 

In this study, we focus on undirected and unweighted networks. However, our methodology can be easily extended and applied on any network type for which graphlets have been defined, such as for directed networks \cite{sarajlic2016graphlet}, for weighted / probabilistic networks \cite{doria2020probabilistic}, hyper-networks \cite{gaudelet2018higher} and for temporal networks.\cite{hulovatyy2015exploring} In the first section of the results, we observe that for node and edge homophily measures, the most homophilic network representations are GPMI and $G_{Adj}$, while for GSI the most homophilic network representations are DeepGraphlets. This disagreement highlights the lack of a single universal measure of homophily that could act as a proxy for selecting the most suitable ML method (linear or nonlinear) for downstream analysis tasks, that would ease the computational burden. This is expected, since we are dealing with computationally intractable (NP-hard) problems, so it is the structure of the data that has to be exploited to construct mining algorithms that are efficient for the data of that particular structure.  From the theory of computation we know that this is the best we can do for computationally intractable problems. Our study is the first to provide fundamental insights into the structural characteristics of network data that enable their \textbf{linear} mining and exploitation, hence providing the building blocks that enable the ML community to build upon them to efficiently and explainably mine complex network data. 

\newpage

\subsection*{Code Availability:} All the scripts used to generate the networks, perform the experiments, and analyze the data are
coded in Python (v3.9.6) and require NumPy, Pandas, SciPy, scikit-learn, NetworkX and MatplotLib. The scripts are available online at \url{https://gitlab.bsc.es/axenos/linear-separability}\\

\subsection*{Data Availability:} The datasets generated and analysed during the current study are available online at \url{https://figshare.com/articles/dataset/Data_zip/26413624}\\

\subsection*{Acknowledgments}
This work is supported by the European Research Council (ERC) Consolidator Grant 770827, the Spanish State Research Agency and the Ministry of Science and Innovation MCIN grant PID2022-141920NB-I00 / AEI /10.13039/501100011033/ FEDER, UE, and the Department of Research and Universities of the Generalitat de Catalunya code 2021 SGR 01536.

\subsection*{Author contributions statement}

A.X. conducted the experiments and wrote the manuscript. N.M.-D. and N.P. conceived and directed the study and contributed to writing of the manuscript. All the authors analyzed the results and reviewed the manuscript.

\subsection*{Competing interests}

The authors declare no competing interests.

\subsection*{Additional information}

\noindent Correspondence should be addressed to natasha@bsc.es

\newpage
\section*{Figures}

\section*{Tables}

\begin{table}[H]
\begin{center}
\begin{tabular}{c | c c c | c c c | c c c |}
& \multicolumn{3}{ c |}{DeepGraphlets} & \multicolumn{3}{ c |}{GPMI} & \multicolumn{3}{ c }{Graphlet Adjacency} \\ 
Networks & L-SVM & SVM RBF & RF & L-SVM  & SVM RBF & RF & L-SVM & SVM RBF & RF \\ \hline
\rowcolor{excelgreen} Cora          & 0.825    & 0.821   & \textbf{0.836} & 0.767    & 0.756   & 0.806 & 0.721    & 0.756   & 0.783 \\ 
\rowcolor{excelgreen} Wikipedia CS      & \textbf{0.821}    & 0.820   & 0.818 & 0.776    & 0.771   & 0.778 & 0.703    & 0.693   & 0.763 \\ 
\rowcolor{excelgreen} CS Co-author  & 0.901    & 0.894   & \textbf{0.906} & 0.849    & 0.833   & 0.864 & 0.809    & 0.783   & 0.853 \\ 
\rowcolor{lightgray} USA air-traffic & 0.655    & 0.637   & 0.668 & 0.670    & 0.671   & \textbf{0.702} & 0.647    & 0.642   & 0.696 \\ 
\rowcolor{lightgray} Chameleon          & 0.679    & 0.641   & \textbf{0.768} & 0.699    & 0.670   & 0.754 & 0.691    & 0.648   & 0.765 \\ 
\rowcolor{lightgray} CiteSeer      & 0.702    & 0.712   & \textbf{0.736} & 0.637    & 0.662   & 0.725 & 0.544    & 0.561   & 0.712 \\ 
\rowcolor{lightgray} Squirrel      & 0.498    & 0.391   & 0.748 & 0.598    & 0.472   & 0.758 & 0.573    & 0.439   & \textbf{0.765} \\ \hline
\rowcolor{lightgreen} Pombe COEX & 0.527 & 0.525 & 0.511 & 0.527 & \textbf{0.53} & 0.503 & 0.518 & 0.519 & 0.498 \\
\rowcolor{lightgreen} Cerevisiae COEX & \textbf{0.611}  & 0.595 & 0.582 & 0.603 & 0.591 & 0.578 & 0.597 & 0.575 & 0.576 \\
\rowcolor{lightgreen} Homo sapiens COEX & 0.507 & \textbf{0.52}  & 0.496 & 0.515 & 0.518 & 0.494 & 0.508 & 0.507 & 0.494 \\
\rowcolor{lightgreen} Pombe PPI  & \textbf{0.671}  & 0.656 & 0.624 & 0.598 & 0.591 & 0.6 & 0.554 & 0.531 & 0.587 \\
\rowcolor{lightgreen} Cerevisiae PPI  & \textbf{0.648}  & 0.633 & 0.627 & 0.633 & 0.624 & 0.615 & 0.607 & 0.596 & 0.599  \\ 
\rowcolor{lightgreen} Homo sapiens PPI  & \textbf{0.519}    & 0.513   & 0.489 & 0.514    & 0.509   & 0.480 & 0.50    & 0.494   & 0.475 \\ \hline
Average & 0.659 & 0.643 & 0.678 & 0.645 & 0.631 & 0.666 & 0.613 & 0.596 & 0.659 \\ \hline
\end{tabular}
\caption[Node classification F1-score in the multi-label molecular networks.]{{\bf Node classification F1-scores in the single-labeled and multi-labeled networks.} For each network (row), the table shows the maximum weighted node classification F1-score of the corresponding classifier (linear SVM (L-SVM), SVM RBF and RF) in the embedding spaces obtained from different matrix representations (columns). The first 7 networks correspond to the single-labeled networks and the last 6 networks to the multi-labeled molecular networks. Note that bold cells indicate the highest value per row. The color-coded scheme presents the level of linear separability in the network's embedding space: green denotes linearly separable embedding space, light green denotes sufficiently linearly separable embedding space, and gray denotes non-linearly separable embedding space.\label{table:all_networks_node_classification}}
\end{center}
\end{table}

\begin{table}[H]
\begin{center}
\begin{tabular}{c|ccc}
Pearson's Correlation                                  &  Synthetic data & Single-label networks  & Multi-label networks  \\ \hline
Node homophily --- F1-score linear SVM                   & \textbf{0.19}  ($1.16 \times 10^{-11}$) & \textbf{0.36} ($1.86 \times 10^{-06}$) & \textbf{0.53} ($7.33 \times 10^{-13}$)\\ \hline
Edge homophily --- F1-score linear SVM                     & \textbf{0.19} ($6.43 \times 10^{-10}$) & \textbf{0.35} ($4.16 \times 10^{-06}$) & \textbf{0.44} ($3.84 \times 10^{-09}$) \\ \hline
GSI --- F1-score linear SVM                               & \textbf{0.68}  ($5.21 \times 10^{-151}$)  & \textbf{0.44}  ($3.17 \times 10^{-10}$) & \textbf{0.46}  ($1.8 \times 10^{-09}$)\\ \hline
\end{tabular}
\caption{{\bf Correlation between homophily levels of the different graphlet-based network representations and the linear separability in the resulting embedding spaces.}  The table shows the Pearson's correlation coefficients between the homophily indexes and the node classification F1-scores obtained by using the linear SVM (Euclidean kernel) in the embedding space of simulated networks from the random partition model (Column 2), of the seven single-labeled networks (Column 3) and of the six molecular multi-labeled networks (Column 4). The values in  parenthesis are the p-values and the bold cells indicate statistical significance of the Pearson's correlation.\label{table:random_partition_model}}
\end{center}
\end{table}

\begin{table}[H]
\begin{center}
\begin{tabular}{c|cc}
Pearson's Correlation & Single-label networks & Multi-label networks \\ \hline
Node homophily - downstream analysis results & \textbf{0.61} ($1.01 \times 10^{-18}$) & \textbf{0.49} ($3.36 \times 10^{-11}$) \\ \hline
Edge homophily - downstream analysis results & \textbf{0.57} ($1.07 \times 10^{-15}$) & \textbf{0.36} ($2.71 \times 10^{-06}$) \\ \hline
GSI - downstream analysis results & \textbf{0.17} ($0.02$) & \textbf{0.36} ($2.56 \times 10^{-06}$)\\ \hline
\end{tabular}
\caption{{\bf Correlation between homophily levels of the different graphlet-based network representations and the results of the downstream analysis tasks.} The table shows the Pearson's correlation coefficients between the homophily indexes and the results of the downstream analysis tasks in the embedding spaces of the seven single-labeled networks (Column 2) and of the six multi-labeled molecular networks (Column 3). In the single-labeled networks, the downstream analysis task is the information retrieval by using the cosine similarity; in the multi-labeled molecular networks, it is the functional module discovery. The values in  parenthesis are the p-values and the bold cells indicate statistical significance of the Pearson's correlation.\label{table:downstream_analysis_correlations}}
\end{center}
\end{table}

\newpage

\bibliographystyle{unsrt}  
\bibliography{sample}  

\newpage

\section{Appendix / supplemental material}

\subsection{Supplementary figures}

\setcounter{figure}{0}

\begin{figure}[H]
    \centering
    \includegraphics[scale=0.11]{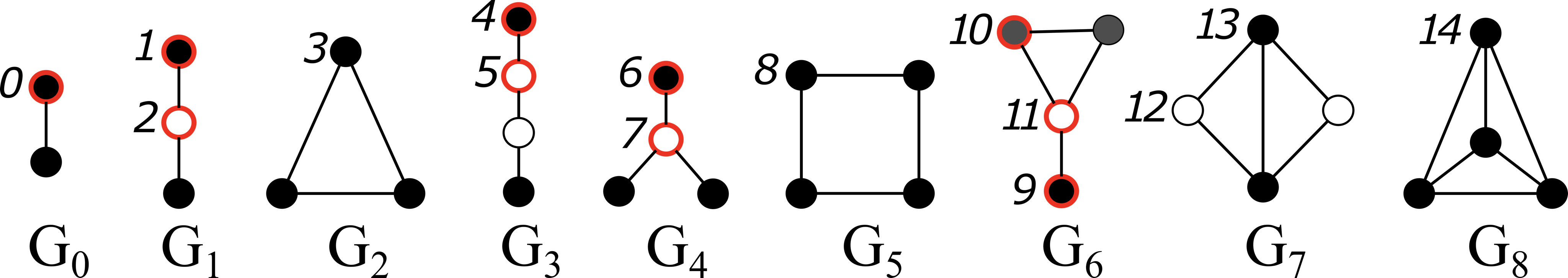}
    \caption[graphlets]{\textbf{The nine 2- to 4-node graphlets and their 15 orbits.} Within each graphlet, $G_i, i \in \{0,\ldots,8\}$, nodes belonging to the same orbit are of the same shade and are numbered from 0 to 14. The eleven non-redundant orbits, whose counts cannot be derived from the counts of the other orbits, are highlighted in red.\cite{Yaveroglu2014}}
    \label{fig:graphlets}
\end{figure}

\begin{figure}[H]
	\begin{center}
		\begin{tabular}{c c}
		\includegraphics[width=8cm]{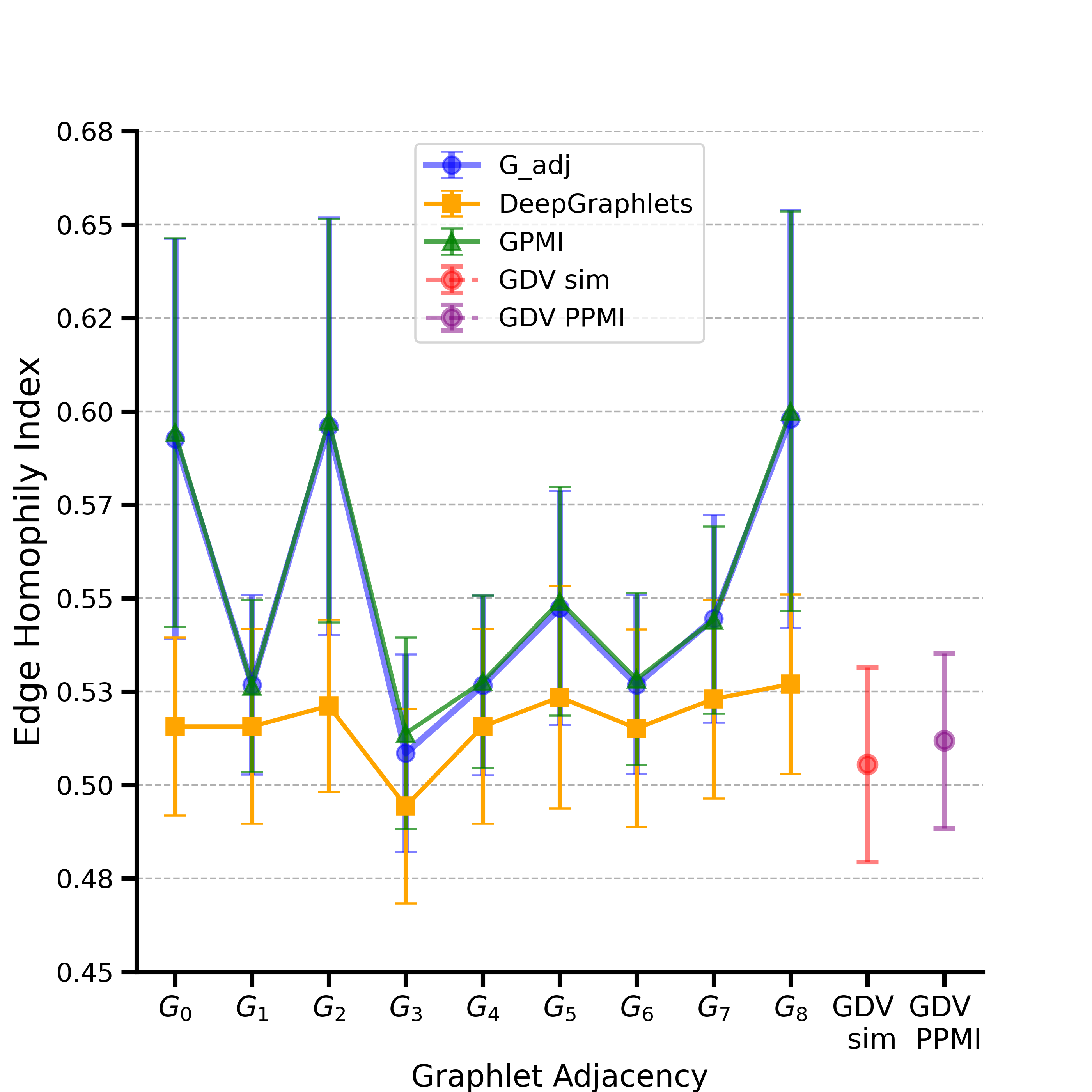} &
        \includegraphics[width=8cm]{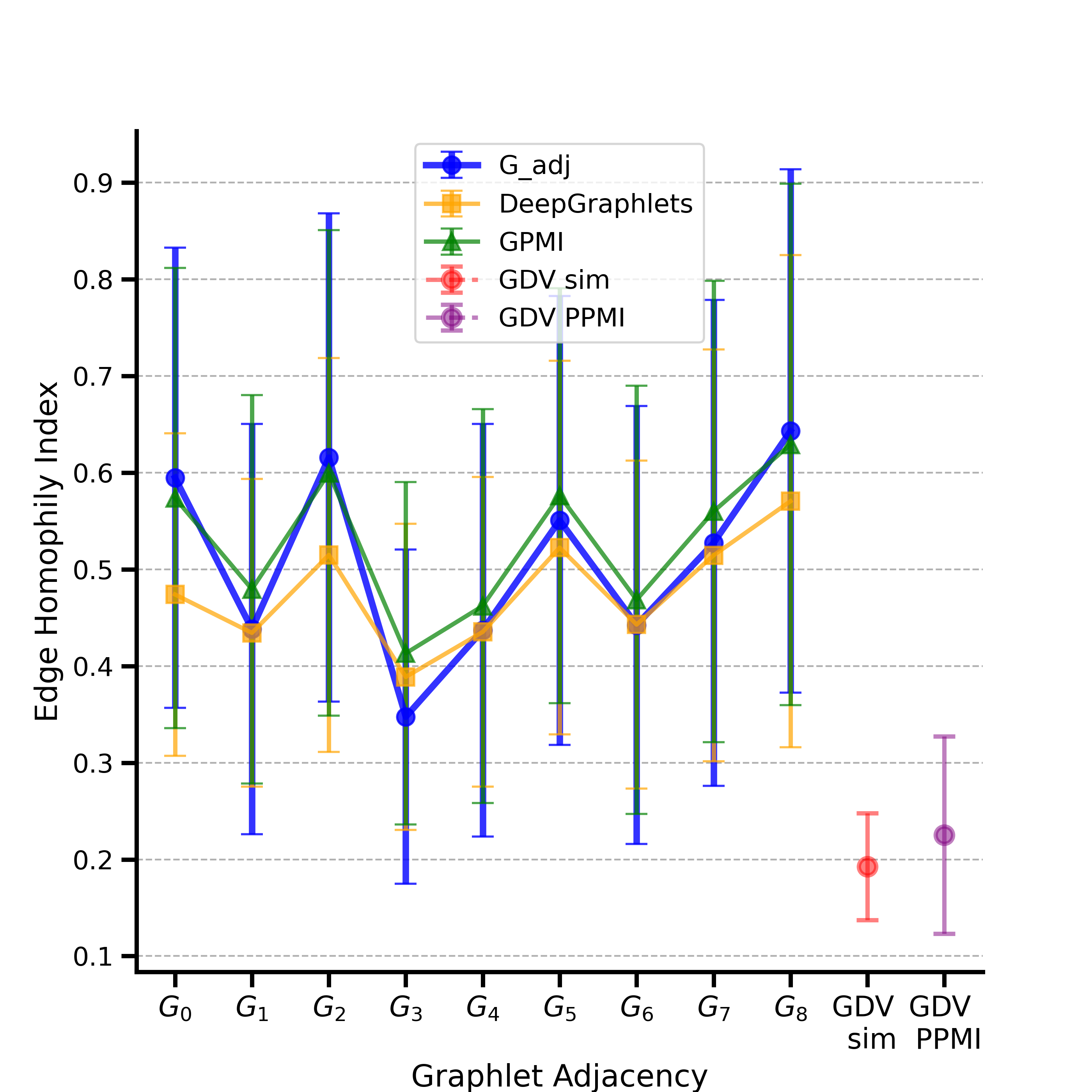} \\
		\end{tabular}
	\end{center}
    \caption{{\bf Graphlet-based network representations are more homophilic.} In the left panel, for each graphlet (x-axis) and for each method (color-coded), the line plot shows on average over the six biological multi-labeled networks the edge homophily index and the standard deviation. The right panel shows the same, but for the seven single-labeled networks. \label{sfig:avg_edge_homophily}}
\end{figure}    

\begin{figure}[H]
    \centering
    \includegraphics[scale=0.45]{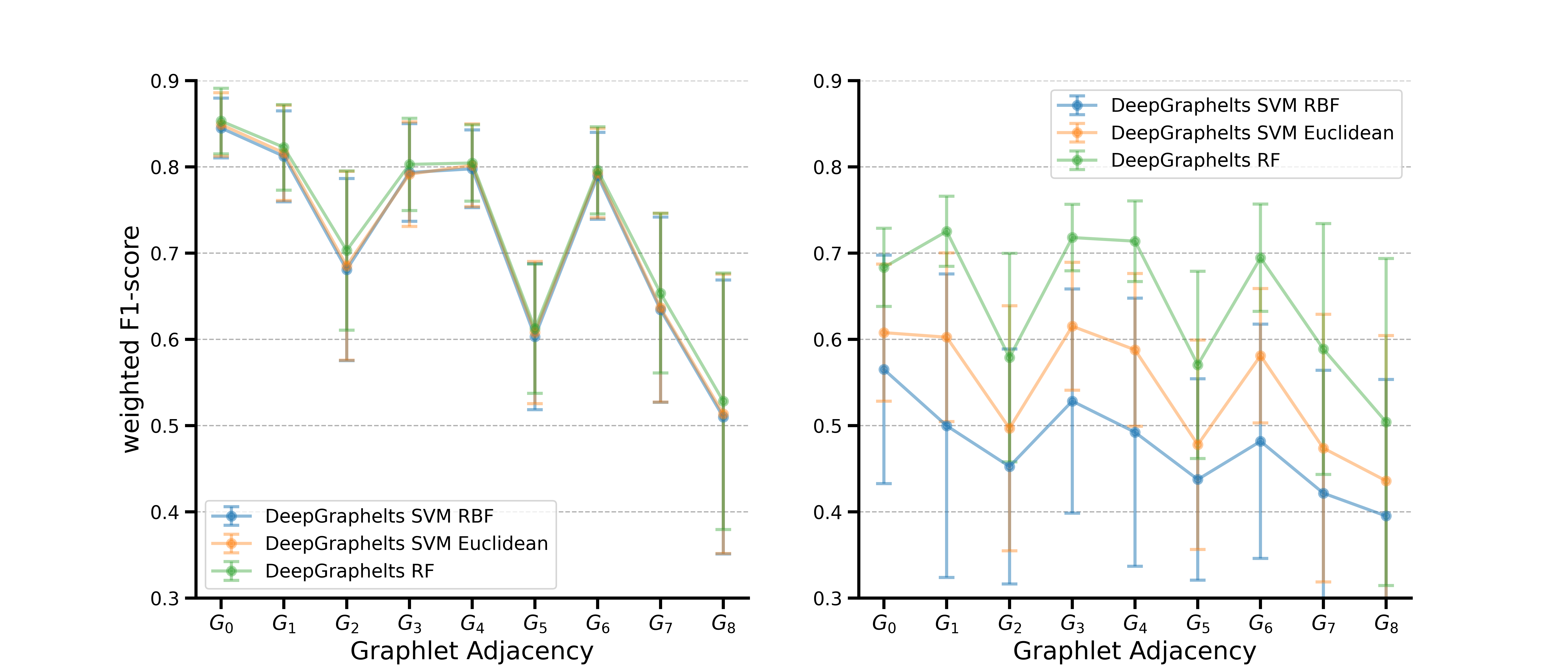}
        \caption{{\bf DeepGraphlets node classification performances in the single-labeled networks.} In the left panel, for each of the nine graphlets (x-axis) and for each classifier (color-coded), the line plot shows for the corresponding nine DeepGraphlets based embeddings the weighted node classification F1-score averaged over the three fully linear single-labeled networks (Cora, CS Co-author and Wikipedia CS) and the standard deviation. The right panel shows the same, but for the four non-linearly separable single-labeled networks (Cameleon, Squirrel, CiteSeer and USA air-traffic).
    \label{sfig:node_classification}}
\end{figure}

\begin{figure}[H]
	\centering
        \includegraphics[scale=0.45]{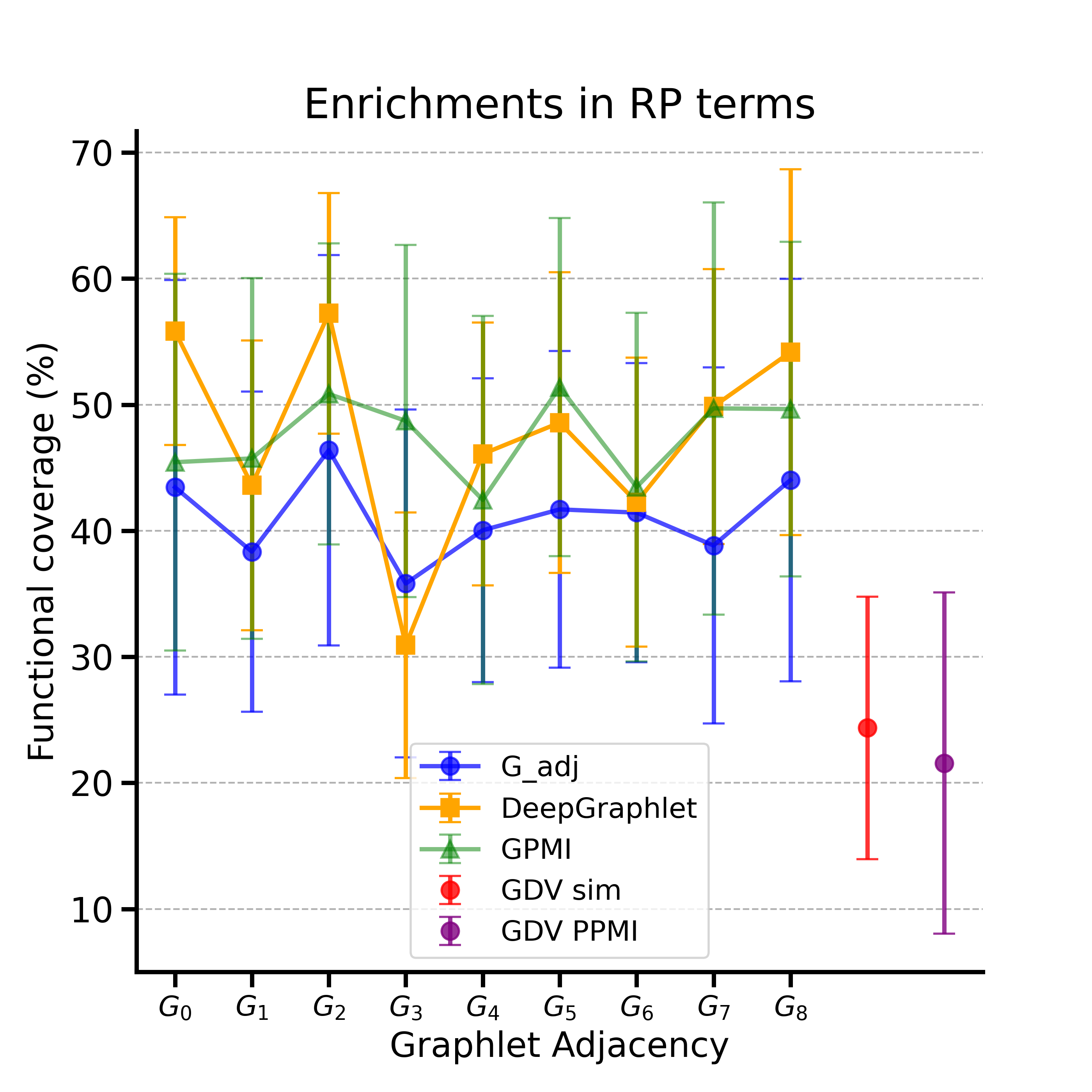} 
    \caption{{\bf Graphlet-based embeddings lead to better functional module discovery in gene embedding spaces for Reactome Pathway (RP) annotations.} For each graphlet (x-axis) and for each method (color-coded), the line plot shows on average (over the six multi-labeled biological networks) percentage of the enriched Reactome Pathway (RP) annotations (y-axis) and the standard deviation. \label{sfig:geneenri}}
\end{figure}

\begin{figure}[H]
	\begin{center}
		\begin{tabular}{c c}
		\hspace{-6.5cm}\textbf{(A)} & \hspace{-6.5cm}\textbf{(B)} \\
        \includegraphics[width=8cm]{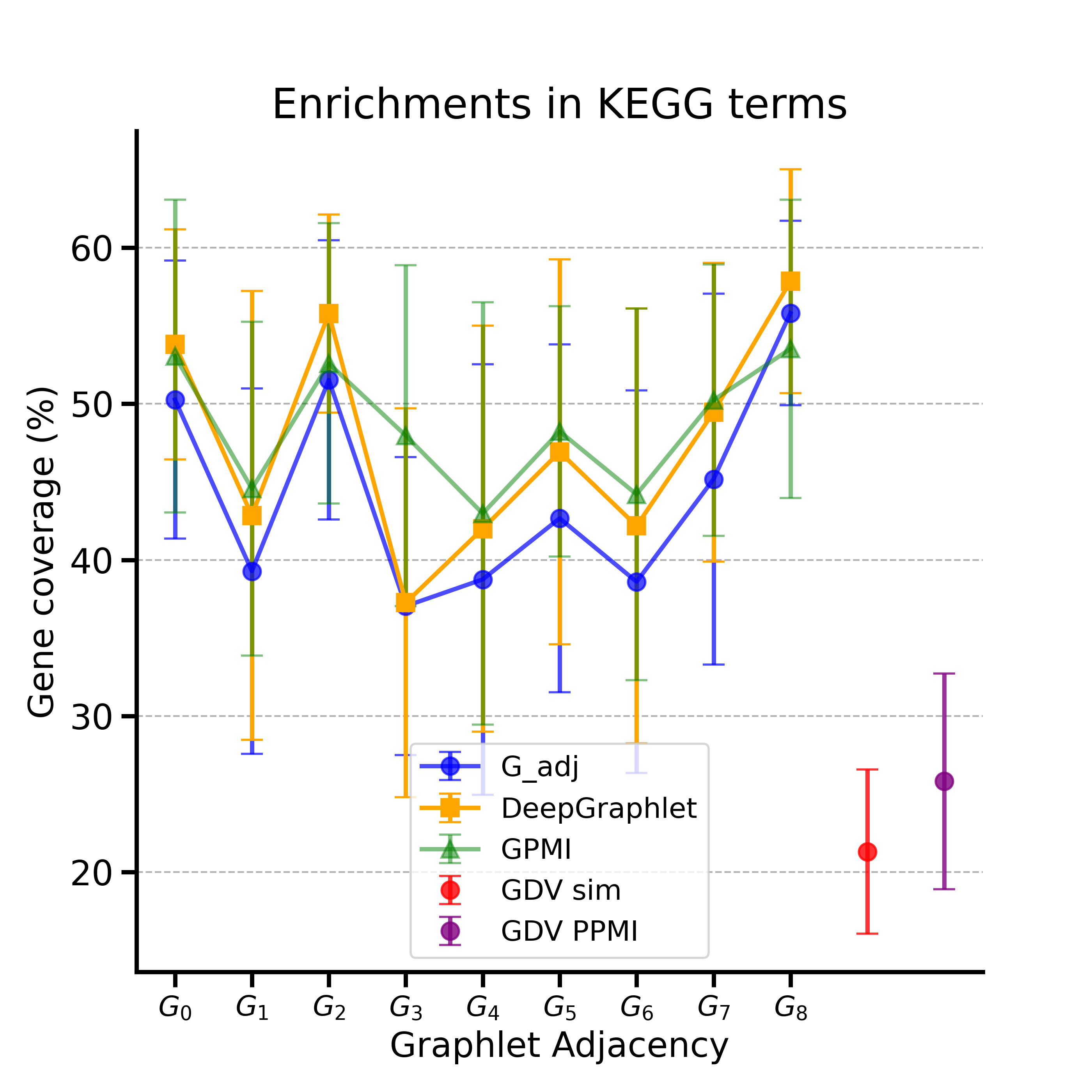} &
        \includegraphics[width=8cm]{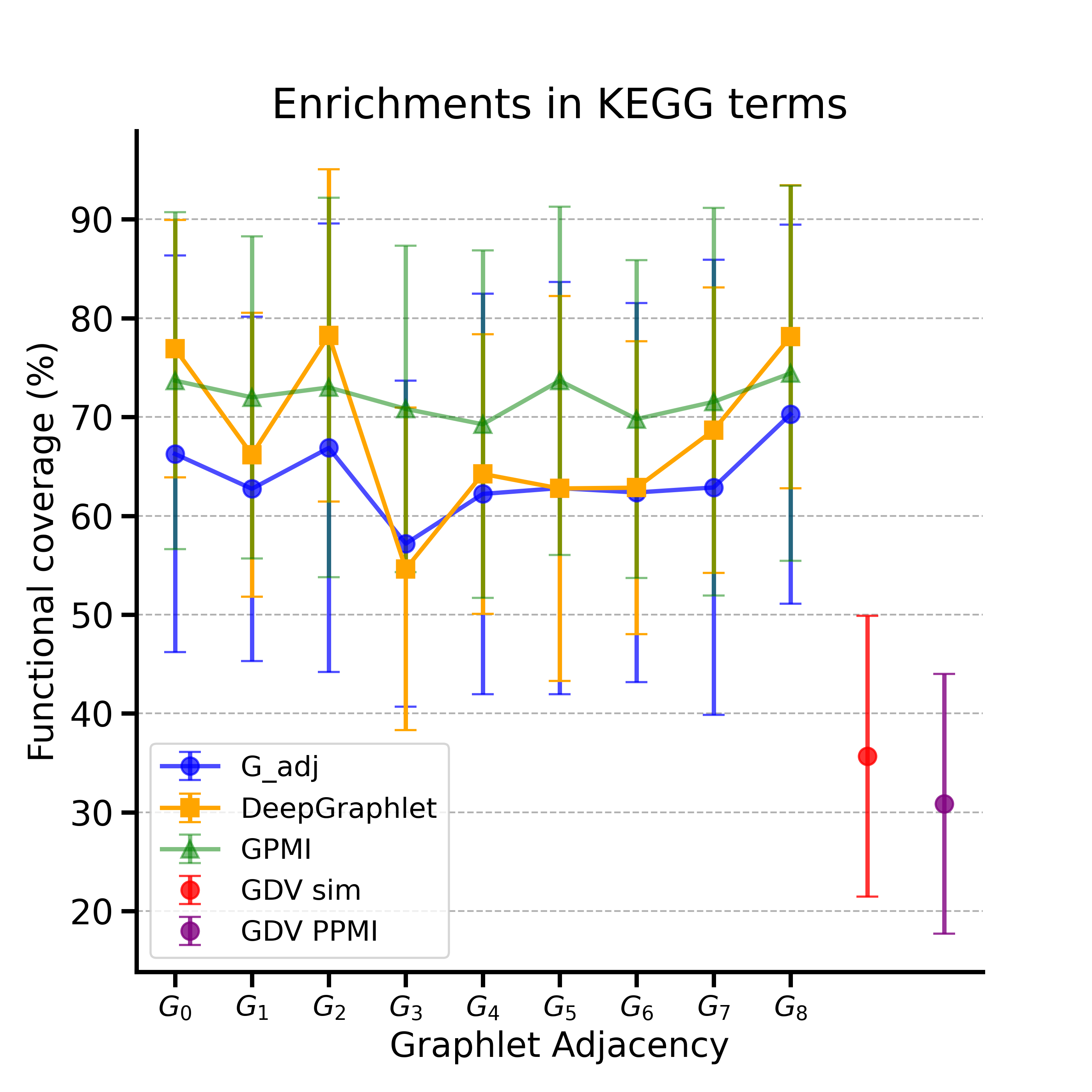} \\
		 \\
		\end{tabular}
	\end{center}
    \caption{{\bf Graphlet-based embeddings lead to better results in downstream analysis tasks for KEGG Pathway annotations.} Panel {\bf A} presents the results of the functional module discovery in gene embedding spaces of the biological multi-labeled networks. In particular, for each graphlet (x-axis) and for each method (color-coded) the line plot shows the average percentage, over the six molecular networks, of annotated genes in the clusters that have at least one KEGG Pathway term enriched in their clusters, along with the standard deviation (y-axis). Panel {\bf B}  shows on average (over the six multi-labeled biological networks)  percentage of the enriched KEGG Pathway annotations (y-axis) and the standard deviation.\label{sfig:Kegg_downstream_analysis_tasks}}
\end{figure}

\begin{figure}[H]
	\begin{center}
		\begin{tabular}{c c}
		\hspace{-6.5cm}\textbf{(A)} & \hspace{-6.5cm}\textbf{(B)} \\
        \includegraphics[width=8cm]{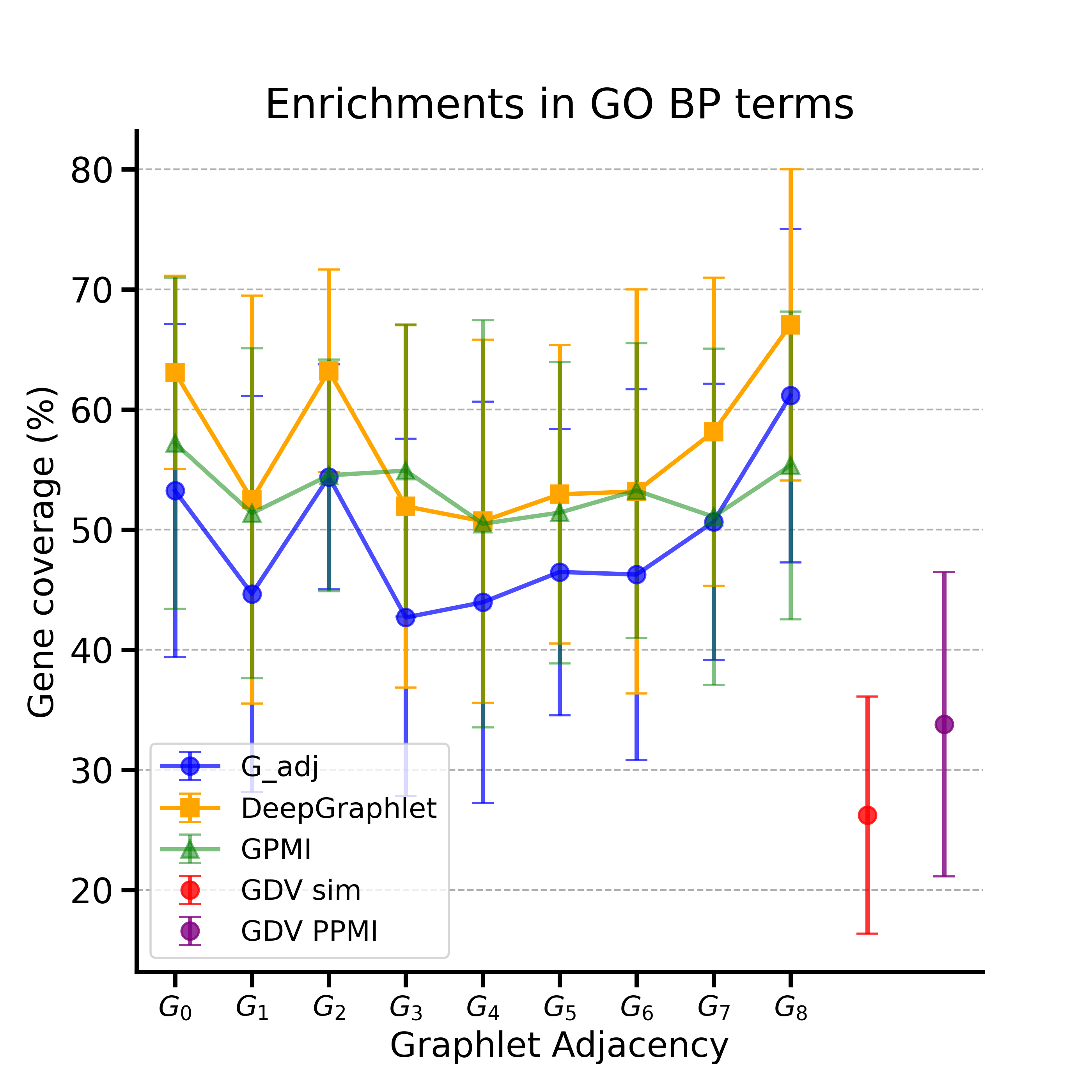} &
        \includegraphics[width=8cm]{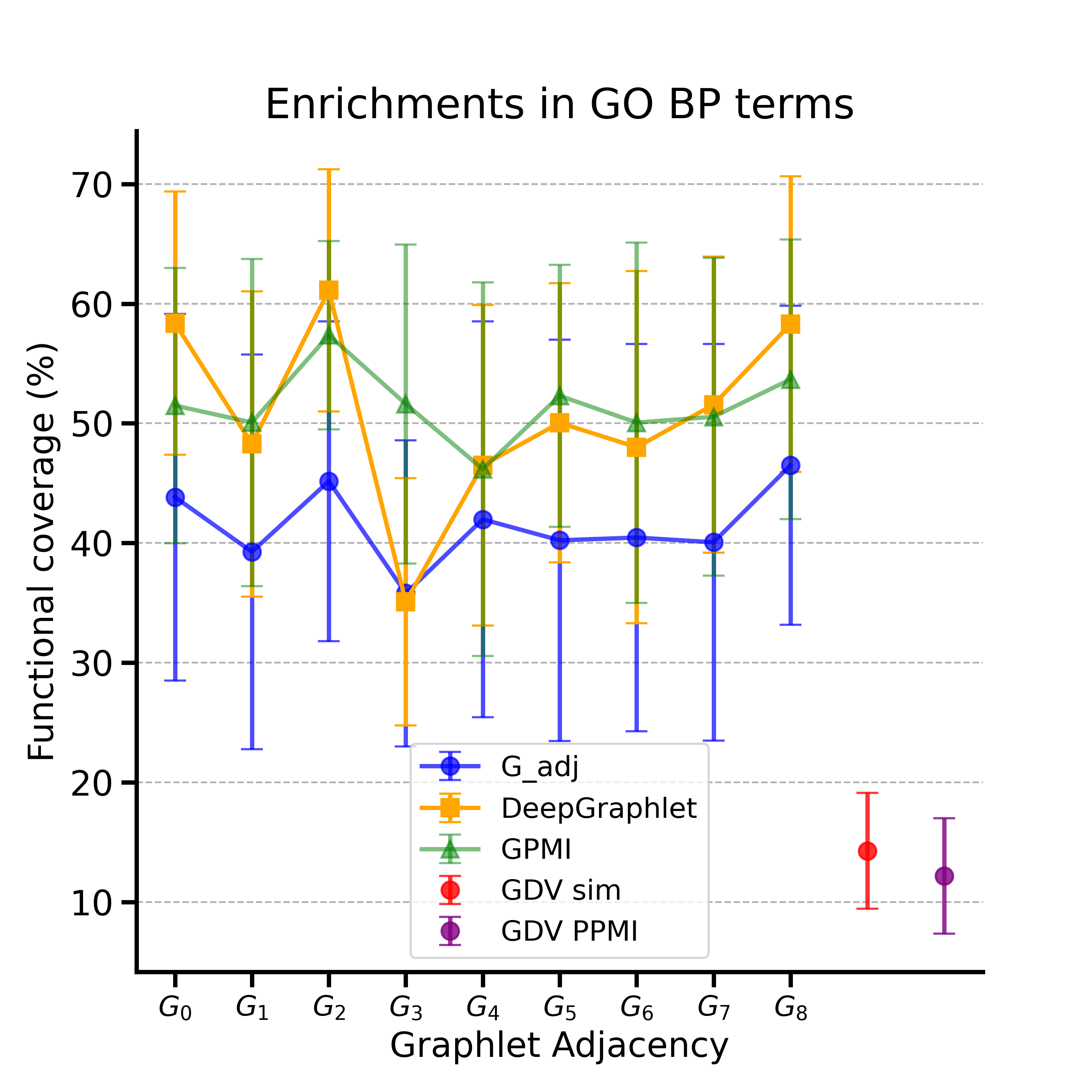} \\
		 \\
		\end{tabular}
	\end{center}
    \caption{{\bf Graphlet-based embeddings lead to better results in downstream analysis tasks for GO BP annotations.} Panel {\bf A} presents the results of the functional module discovery in gene embedding spaces of the biological multi-labeled networks. In particular, for each graphlet (x-axis) and for each method (color-coded) the line plot shows the average percentage, over the six molecular networks, of annotated genes in the clusters that have at least one GO BP term enriched in their clusters, along with the standard deviation (y-axis). Panel {\bf B}  shows on average (over the six multi-labeled biological networks) percentage of the enriched GO BP annotations (y-axis) and the standard deviation.\label{sfig:go_bp_downstream_analysis_tasks}}
\end{figure}

\subsection{Supplementary tables}

\setcounter{table}{0}

\begin{table}[H]
\begin{center}
\begin{tabular}{c | c c c | c c c |}
& \multicolumn{3}{ c |}{PPI} & \multicolumn{3}{ c |}{COEX} \\
Species & \#Node & \#Edges & Density & \#Node & \#Edges & Density \\
\hline
\begin{tabular}[c]{@{}c@{}}\emph{Homo sapiens} \\ (Human)\end{tabular}
& 18,614 & 398,713 & 0.0023 & 16,808 & 1,610,160 & 0.0114 \\
\hline
\begin{tabular}[c]{@{}c@{}}\emph{Saccharomyces cerevisiae} \\ (Baker's yeast)\end{tabular}
& 5,886 & 110,451 & 0.0064 & 5,647 & 167,560 & 0.0105 \\
\hline
\begin{tabular}[c]{@{}c@{}}\emph{Schizosaccharomyces pombe} \\ (Fission yeast)\end{tabular}
& 3,214 & 10,923 & 0.0021 & 4,951 & 504,790 & 0.0412 \\
\hline
\end{tabular}
\caption{{\bf Statistics of the multi-labeled biological networks.} For each species (row), the table shows the corresponding number of nodes, edges, and density for the PPI (columns 2, 3, and 4) and COEX (columns 5, 6, and 7) networks. \label{table:bio_netstats}}
\end{center}
\end{table}

\begin{table}[H]
\begin{center}
\begin{tabular}{ccccccc}
Network         & Labels (categories) & \#Nodes & \#Edges & Density & Node Homophily & Edge Homophily \\ \hline
Wikipedia CS    & 10                  & 10,657  & 145,133 & 0.0025  & 0.695          & 0.684          \\ \hline
Cameleon        & 5                   & 2,277   & 31,371  & 0.0121  & 0.247          & 0.229          \\ \hline
Squirrel        & 5                   & 5,201   & 188,239 & 0.0139  & 0.217          & 0.222          \\ \hline
Cora            & 7                   & 2,708   & 5,278   & 0.0014  & 0.825          & 0.809          \\ \hline
CiteSeer        & 6                   & 2,120   & 4,732   & 0.0012  & 0.711          & 0.735          \\ \hline
USA air-traffic & 4                   & 1,186   & 13,597  & 0.0193  & 0.371          & 0.667          \\ \hline
CS Co-author    & 15                  & 18,333  & 81,894  & 0.0005  & 0.832          & 0.684          \\ \hline
\end{tabular}
\caption{{\bf Statistics of the single-labeled networks.} For each network (row), the table shows the corresponding number of labels (column 2), nodes (column 3), edges (column 4) and density (column 5). In addition, it shows the node homophily index (column 6) and the edge homophily index (column 7) computed on their adjacency matrix representations. \label{table:netstats}}
\end{center}
\end{table}

\begin{table}[H]
\begin{center}
\begin{tabular}{c | c | c c | c c |}
& GO BP & \multicolumn{2}{ c |}{PPI} & \multicolumn{2}{ c |}{COEX} \\
& Categories  & \#Annotated Nodes & \#Nodes & \#Annotated Nodes & \#Nodes \\ \hline
\begin{tabular}[c]{@{}c@{}}\emph{Homo sapiens} \\ (Human)\end{tabular}
& 18 & 8,772 & 18,614 & 7,775 & 16,808 \\ \hline
\begin{tabular}[c]{@{}c@{}}\emph{Saccharomyces cerevisiae} \\ (Baker's yeast)\end{tabular}
& 12 & 4,591 & 5,886 & 4,517 & 5,647 \\ \hline
\begin{tabular}[c]{@{}c@{}}\emph{Schizosaccharomyces pombe} \\ (Fission yeast)\end{tabular}
& 11 & 1,413 & 3,214 & 1,669 & 4,951 \\ \hline
\end{tabular}
\caption{{\bf Numbers of annotated nodes in the biological multi-labeled networks.} For each species (row), the table shows the numbers of level 1 GO BP terms that annotate the nodes (column 2), the numbers of nodes in the PPI networks annotated with GO BP terms and the total numbers of nodes in the PPI networks (columns 3 and 4), and the numbers of nodes in the COEX networks annotated with GO BP terms and the total numbers of nodes in the COEX networks (columns 5 and 6).  \label{stable:bio_annotated_nodes}}
\end{center}
\end{table}

\begin{table}[H]
\begin{center}
\begin{tabular}{c | c c | c c |}
& \multicolumn{2}{ c |}{PPI} & \multicolumn{2}{ c |}{COEX} \\
& \#Annotated Nodes & \#Nodes & \#Annotated Nodes & \#Nodes \\ \hline
\begin{tabular}[c]{@{}c@{}}\emph{Homo sapiens} \\ (Human)\end{tabular}
& 7,770 & 18,614 & 6,836 & 16,808 \\ \hline
\begin{tabular}[c]{@{}c@{}}\emph{Saccharomyces cerevisiae} \\ (Baker's yeast)\end{tabular}
& 2,131 & 5,886 & 2,109 & 5,647 \\ \hline
\begin{tabular}[c]{@{}c@{}}\emph{Schizosaccharomyces pombe} \\ (Fission yeast)\end{tabular}
& 1,298 & 3,214 & 1,712 & 4,951 \\ \hline
\end{tabular}
\caption{{\bf Numbers of annotated nodes in the biological multi-labeled networks with KEGG pathway terms.} For each species (row), the table shows the numbers of nodes in the PPI networks annotated with KEGG pathway terms and the total numbers of nodes in the PPI networks (columns 2 and 3), and the numbers of nodes in the COEX networks annotated with KEGG terms and the total numbers of nodes in the COEX networks (columns 4 and 5).  \label{stable:kegg_annotated_nodes}}
\end{center}
\end{table}

\begin{table}[H]
\begin{center}
\begin{tabular}{c | c c | c c |}
& \multicolumn{2}{ c |}{PPI} & \multicolumn{2}{ c |}{COEX} \\
& \#Annotated Nodes & \#Nodes & \#Annotated Nodes & \#Nodes \\ \hline
\begin{tabular}[c]{@{}c@{}}\emph{Homo sapiens} \\ (Human)\end{tabular}
& 9,951 & 18,614 & 8,934 & 16,808 \\ \hline
\begin{tabular}[c]{@{}c@{}}\emph{Saccharomyces cerevisiae} \\ (Baker's yeast)\end{tabular}
& 1,699 & 5,886 & 1,688 & 5,647 \\ \hline
\begin{tabular}[c]{@{}c@{}}\emph{Schizosaccharomyces pombe} \\ (Fission yeast)\end{tabular}
& 1,163 & 3,214 & 1,449 & 4,951 \\ \hline
\end{tabular}
\caption{{\bf Numbers of annotated nodes in the biological multi-labeled networks with Reactome Pathway terms.} For each species (row), the table shows the numbers of nodes in the PPI networks annotated with RP terms and the total numbers of nodes in the PPI networks (columns 2 and 3), and the numbers of nodes in the COEX networks annotated with RP terms and the total numbers of nodes in the COEX networks (columns 4 and 5).  \label{stable:rp_annotated_nodes}}
\end{center}
\end{table}

\begin{table}[H]
\begin{center}
\begin{tabular}{c | c c c | c c c | c c c |}
& \multicolumn{3}{ c |}{DeepGraphlets} & \multicolumn{3}{ c |}{GPMI} & \multicolumn{3}{ c }{Graphlet Adjacency} \\ 
Networks & L-SVM & SVM RBF & RF & L-SVM  & SVM RBF & RF & L-SVM & SVM RBF & RF \\ \hline
\rowcolor{excelgreen} Cora          & $G_0$ & $G_0$ & $\textbf G_0$ & $G_3$ & $G_3$ & $G_1$ & $G_3$ & $G_3$ & $G_3$ \\ 
\rowcolor{excelgreen} Wikipedia CS   & $\textbf G_0$ & $G_0$ & $G_0$ & $G_3$ & $G_3$ & $G_0$ & $G_4$ & $G_4$ & $G_0$ \\ 
\rowcolor{excelgreen} CS Co-author   & $G_0$ & $G_0$ & $\textbf G_0$ & $G_3$ & $G_3$ & $G_1$ & $G_3$ & $G_3$ & $G_0$ \\ 
\rowcolor{lightgray} USA air-traffic & $G_1$ & $G_6$ & $G_6$ & $G_7$ & $G_2$ & $\textbf G_2$ & $G_6$ & $G_1$ & $G_0$ \\ 
\rowcolor{lightgray} Chameleon       & $G_0$ & $G_0$ & $\textbf G_4$ & $G_1$ & $G_1$ & $G_4$ & $G_6$ & $G_6$ & $G_1$ \\ 
\rowcolor{lightgray} CiteSeer        & $G_3$ & $G_1$ & $\textbf G_1$ & $G_3$ & $G_3$ & $G_1$ & $G_3$ & $G_3$ & $G_3$ \\ 
\rowcolor{lightgray} Squirrel        & $G_3$ & $G_3$ & $ G_4$ & $G_4$ & $G_1$ & $G_4$ & $G_1$ & $G_1$ & $\textbf G_1$ \\ \hline
\rowcolor{lightgreen} Pombe COEX & $G_2$ & $G_4$ & $G_0$ & $G_7$ & $\textbf G_3$ & $G_3$ & $G_2$ & $G_4$ & $G_7$ \\
\rowcolor{lightgreen} Cerevisiae COEX & $\textbf G_0$ & $G_0$ & $G_0$ & $G_3$ & $G_6$ & $G_3$ & $G_6$ & $G_2$ & $G_5$ \\
\rowcolor{lightgreen} Homo sapiens COEX & $G_8$ & $\textbf G_0$ & $G_2$ & $G_2$ & $G_3$ & $G_7$ & $G_0$ & $G_1$ & $G_7$ \\
\rowcolor{lightgreen} Pombe PPI & $\textbf G_0$ & $G_0$ & $G_0$ & $G_2$ & $G_3$ & $G_7$ & $G_2$ & $G_3$ & $G_2$ \\
\rowcolor{lightgreen} Cerevisiae PPI & $\textbf G_8$ & $G_8$ & $G_8$ & $G_8$ & $G_8$ & $G_8$ & $G_8$ & $G_8$ & $G_8$ \\
\rowcolor{lightgreen} Homo sapiens PPI & $\textbf G_2 $ & $G_2$ & $G_8$ & $G_8$ & $G_2$ & $G_8$ & $G_8$ & $G_0$ & $G_8$ \\ \hline
\end{tabular}
\caption{{\bf Embedding space that corresponds to the maximum node classification F1-scores in the single-labeled and multi-labeled networks.} For each network (row), the table shows the graphlet-based embedding space in which the maximum weighted node classification F1-score of the corresponding classifier (linear SVM (L-SVM), SVM RBF, and RF) is achieved. The first 7 networks correspond to the single-labeled networks, and the last 6 networks correspond to the multi-labeled molecular networks. The color-coded scheme represents the level of linear separability in the network's embedding space: green denotes linear separability, light green denotes sufficient linear separability, and gray denotes non-linear separability. Bold cells denote the graphlet that yielded the highest value per row.\label{stable:all_networks_node_classification_index}}
\end{center}
\end{table}

\begin{table}[H]
\begin{center}
\begin{tabular}{c |ccccccc}
      & Wikipedia CS & Chameleon  & Cora  & USA air-traffic & CS Co-author & Squirrel & CiteSeer \\ \hline
$G_0$ & 100          & 100       & 100   & 100             & 100          & 100      & 100      \\ \hline
$G_1$ & 100          & 100       & 95.2  & 100             & 100          & 100      & 100      \\ \hline
$G_2$ & 80.48        & 91.35     & 54.28 & 79.85           & 86.68        & 91.6     & \textbf{40.38}    \\ \hline
$G_3$ & 100          & 100       & 94.28 & 100             & 100          & 100      & 100      \\ \hline
$G_4$ & 99.41        & 99.47     & 88.99 & 99.66           & 97.83        & 99.98    & 89.01    \\ \hline
$G_5$ & 78.99        & 83.57     & \textbf{41.54} & 64.08         & 60.68        & 94.02    & \textbf{35.9}     \\ \hline
$G_6$ & 99.08        & 98.68     & 83.64 & 99.49           & 98.06        & 99.37    & 73.16    \\ \hline
$G_7$ & 78.91        & 90.38     & \textbf{43.57} & 78.75           & 77.79        & 91.12    & \textbf{29.58}    \\ \hline
$G_8$ & 61.64        & 78.92     & \textbf{14.51} & 63.58           & 61.86        & 80.02    & \textbf{8.77}     \\ \hline
\end{tabular}

\caption{{\bf Graphlet coverage of the single-labeled networks.} For each network (columns 2-8), the table shows the percentage of nodes that touch (participate in) a given graphlet (row). \label{stable:graphlet_coverage_nets}}
\end{center}
\end{table}

\begin{table}[H]
\begin{center}
\begin{tabular}{c | ccc | ccc |}
& \multicolumn{3}{c|}{PPI} & \multicolumn{3}{c |}{COEX} \\
Graphlet & \begin{tabular}[c]{@{}c@{}}Homo \\ sapiens \end{tabular} & \begin{tabular}[c]{@{}c@{}}Saccharomyces \\ cerevisiae \end{tabular} & \begin{tabular}[c]{@{}c@{}}Schizosaccharomyces \\ pombe \end{tabular} & \begin{tabular}[c]{@{}c@{}}Homo \\ sapiens \end{tabular} & \begin{tabular}[c]{@{}c@{}}Saccharomyces \\ cerevisiae \end{tabular} & \begin{tabular}[c]{@{}c@{}}Schizosaccharomyces \\ pombe \end{tabular} \\ \hline
$G_0$ & 100 & 100 & 100 & 100 & 100 & 100 \\ \hline
$G_1$ & 100 & 100 & 100 & 100 & 100 & 100 \\ \hline
$G_2$ & 76.99 & 92.49 &  \textbf{50.22} & 99.96 & 97.91 & 95.64 \\ \hline
$G_3$ & 100 & 100 & 100 & 100 & 100 & 100 \\ \hline
$G_4$ & 99.94 & 100 & 98.88 & 100 & 99.84 & 99.56 \\ \hline
$G_5$ & 86.48 & 92.51 & 55.01 & 99.9 & 97.11 & 94.3 \\ \hline
$G_6$ & 99.77 & 100 & 96.86 & 100 & 99.77 & 99.41 \\ \hline
$G_7$ & 76.6 & 92.46 & \textbf{47.51} & 99.92 & 99.59 & 95.03 \\ \hline
$G_8$ & 56.15 & 82.91 & \textbf{26.38} & 99.57 & 94.1 & 90.06 \\ \hline
\end{tabular}
\caption{{\bf Graphlet coverage for the PPI and COEX multi-labeled molecular networks.} For each species (columns 2-7), the table shows the percentage of nodes that participate in a given graphlet (rows) for both PPI and COEX networks. The bold cells indicate graphlet coverage below 50 \%, so these graphlets are excluded from the analysis.
\label{stable:graphlet_coverage_bio_nets}}
\end{center}
\end{table}

\begin{table}[H]
\begin{center}
\begin{tabular}{c|ccccccc}
DeepGraphlets & Wikipedia CS & Chameleon & Squirrel & USA air-traffic & Cora  & CiteSeer & CS Co-author \\ \hline
$G_0$         & 0.756        & 0.536     & 0.503    & 0.539   & 0.703 & 0.654    & 0.890        \\ \hline
$G_1$         & \textbf{0.766}        & 0.543     & 0.506    & 0.555  & \textbf{0.745} & 0.672   & \textbf{0.896}        \\ \hline
$G_2$         & 0.672        & 0.527     & 0.503    & 0.540           & 0.535 & 0.507    & 0.800        \\ \hline
$G_3$         & 0.689        & \textbf{0.545}     & \textbf{0.512}    & 0.562           & 0.752 & \textbf{0.696}    & 0.881        \\ \hline
$G_4$         & 0.755        & 0.542     & 0.506    & 0.555           & 0.726 & 0.642    & 0.885        \\ \hline
$G_5$         & 0.681        & 0.539     & 0.503    & 0.569           & 0.528 & 0.512    & 0.683        \\ \hline
$G_6$         & 0.759        & 0.541     & 0.507    & 0.546           & 0.703 & 0.591    & 0.881        \\ \hline
$G_7$         & 0.680        & 0.529     & 0.504    & \textbf{0.571}           & 0.524 & 0.507    & 0.758        \\ \hline
$G_8$         & 0.596        & 0.527     & 0.504    & 0.557           & 0.501 & 0.501    & 0.647        \\ \hline
\end{tabular}
\caption[DeepGraphlets weighted AUROC in the single-label networks.]{{\bf DeepGraphlets weighted AUROC in the single-labeled networks.} For each network (column 2-8), the table shows the weighted AUROC of the embedding space that corresponds to a given graphlet-based extension of DeepWalk (rows). The bold cells present the highest value per column.  \label{stable:single_label_networks_deepgraphlets}}
\end{center}
\end{table}

\begin{table}[H]
\begin{center}
\begin{tabular}{c|ccccccc}
   GPMI   & Wikipedia CS & Chameleon & Squirrel & USA air-traffic & Cora  & CiteSeer & CS Co-author \\ \hline
$G_0$ & 0.643        & 0.533     & 0.504    & 0.532          & 0.645 & 0.589    & 0.769        \\ \hline
$G_1$ & 0.718        & 0.536     & 0.508    & 0.540          & 0.663 & 0.614    & \textbf{0.839}       \\ \hline
$G_2$ & 0.597        & 0.532     & 0.504    & 0.536          & 0.530 & 0.506    & 0.700        \\ \hline
$G_3$ & \textbf{0.738} & \textbf{0.542}     & 0.507    & 0.543        & \textbf{0.697} & \textbf{0.644}    & 0.861        \\ \hline
$G_4$ & 0.708        & 0.525     & 0.506    & 0.543           & 0.647 & 0.593    & 0.816        \\ \hline
$G_5$ & 0.625        & 0.525     & 0.505    & 0.541           & 0.522 & 0.509    & 0.623        \\ \hline
$G_6$ & 0.711        & 0.529     & \textbf{0.509}    & 0.539           & 0.629 & 0.556    & 0.824        \\ \hline
$G_7$ & 0.610        & 0.531     & 0.505    & 0.545           & 0.516 & 0.505    & 0.684        \\ \hline
$G_8$ & 0.549        & 0.527     & 0.504    & \textbf{0.555}           & 0.501 & 0.501    & 0.601        \\ \hline
\end{tabular}
\caption[GPMI weighted AUROC in the single-label networks.]{{\bf GPMI weighted AUROC in the single-labeled networks.} For each network (columns 2-8), the table shows the weighted AUROC of the embedding space that corresponds to a given graphlet-based extension of PPMI (rows). The bold cells present the highest value per column.  \label{stable:single_label_networks_gpmi}}
\end{center}
\end{table}

\begin{table}[H]
\begin{center}
\begin{tabular}{c|ccccccc}
   G Adj   & Wikipedia CS & Chameleon & Squirrel & USA air-traffic & Cora  & CiteSeer & CS Co-author \\ \hline
$G_0$ & 0.656        & 0.526     & 0.502    & 0.547           & 0.639 & 0.583    & 0.752        \\ \hline
$G_1$ & 0.696        & 0.529     & 0.507    & 0.560           & 0.698 & 0.634    & 0.829        \\ \hline
$G_2$ & 0.598        & 0.529     & 0.505    & 0.535           & 0.531 & 0.505    & 0.684        \\ \hline
$G_3$ & 0.549        & 0.524     & \textbf{0.508} & 0.560  & \textbf{0.705} & \textbf{0.670}    & \textbf{0.838}        \\ \hline
$G_4$ & \textbf{0.697} & 0.527     & 0.507    & 0.558  & 0.667 & 0.611    & 0.820        \\ \hline
$G_5$ & 0.618        & 0.529     & 0.507    & 0.543  & 0.519 & 0.507    & 0.624        \\ \hline
$G_6$ & 0.684        & 0.528     & 0.507    & \textbf{0.565} & 0.648 & 0.564    & 0.821        \\ \hline
$G_7$ & 0.622        & \textbf{0.532}     & 0.505    & 0.546           & 0.518 & 0.503    & 0.678        \\ \hline
$G_8$ & 0.552        & 0.526     & 0.504    & 0.544           & 0.501 & 0.501    & 0.598        \\ \hline
\end{tabular}
\caption[ Graphlet Adjacency weighted AUROC in the single-label networks.]{{\bf Graphlet Adjacency weighted AUROC in the single-labeled networks.} For each network (column 2-8), the table shows the weighted AUROC of the embedding space that corresponds to a given graphlet-based extension of adjacency matrix (rows). The bold cells present the highest value per column.  \label{stable:single_label_networks_g_adj}}
\end{center}
\end{table}


\end{document}